\documentclass[fleqn,10pt]{wlscirep}
\usepackage[utf8]{inputenc}
\usepackage[T1]{fontenc}

\usepackage{bm}
\usepackage{subcaption}
\graphicspath{{figures/}}

\newcommand{\bs}{\boldsymbol}

\usepackage{lineno}

\title{Robust real-time imaging through flexible multimode fibers}

\author[1,*]{Abdullah Abdulaziz}
\author[2]{Simon Peter Mekhail}
\author[1]{Yoann Altmann}
\author[2]{Miles J. Padgett}
\author[1]{Stephen McLaughlin}
\affil[1]{Heriot-Watt University, School of Engineering and Physical Sciences, Edinburgh, EH14 4AS, United Kingdom}
\affil[2]{University of Glasgow, School of Physics and Astronomy, Glasgow, G12 8QQ, United Kingdom}

\affil[*]{a.abdulaziz@hw.ac.uk}

\begin{abstract}
Conventional endoscopes comprise a bundle of optical fibers, associating one fiber for each pixel in the image. In principle, this can be reduced to a single multimode optical fiber (MMF), the width of a human hair, with one fiber spatial-mode per image pixel. However, images transmitted through a MMF emerge as unrecognizable speckle patterns due to dispersion and coupling between the spatial modes of the fiber. Furthermore, speckle patterns change as the fiber undergoes bending, making the use of MMFs in flexible imaging applications even more complicated. 
In this paper, we propose a real-time imaging system using flexible MMFs, but which is robust to bending. 
Our approach does not require access or feedback signal from the distal end of the fiber during imaging.
We leverage a variational autoencoder (VAE) to reconstruct and classify images from the speckles and show that these images can still be recovered when the bend configuration of the fiber is changed to one that was not part of the training set. We utilize a MMF $300$~mm long with a 62.5~$\mu$m core for imaging $10~\times~10$~cm objects placed approximately at $20$~cm from the fiber and the system can deal with a change in fiber bend of 50$^\circ$ and range of movement of 8~cm.

\end{abstract}
\begin{document}

\flushbottom
\maketitle

\section*{Introduction}
\label{intro}

Multimode fibers (MMFs), the width of a human hair, potentially allow for the transmission of images formed from the thousands of spatial modes they support~\cite{psaltis2016,stellinga2021}. This creates the potential for minimally invasive and high-resolution imaging systems such as ultra-thin endoscopes which can be used for imaging objects out of the reach of conventional technology. Unfortunately, even when using temporally coherent light, images propagated through MMFs suffer from severe spatial distortions and can appear as random speckle patterns at the distal end due to modal dispersion in the fiber~\cite{psaltis2016,stellinga2021}. However, although the information is completely scrambled at either end of a MMF, the vast majority of the information is not lost and hence the image can, in principle, be recovered.

To image through MMFs, two main classes of methods have been used to date. The first consists of raster-scanning methods~\cite{cizmar2012,bianchi2012,ploschner2015,loterie2015,caravaca2017,stellinga2021}, which rely on measuring the complex mapping of the input field onto the output field, namely the transmission matrix (TM)~\cite{popoff2010,cizmar2011,carpenter2014,n2018,li2021}. Using the TM, the input field at the proximal facet of a MMF can be specified such that focused light beams are generated at the distal end. By calculating the correct series of input fields, this focused beam can be scanned over an object with a field of view defined by the numerical aperture of the fiber. The second class of methods consists of speckle imaging approaches~\cite{choi2012,amitonova2018,caravaca2019,lan2019,lan2020}. With these methods, a set of speckle patterns is recorded at the distal end of a MMF, forming a measurement matrix during a calibration stage. During imaging, the same speckle patterns are projected sequentially onto a new object and the collected overall signal provides a speckle measurement. Iterative methods have been traditionally used to reconstruct objects from the speckle measurements and the measurement matrix. Although Lan \textit{et al.}~\cite{lan2020} showed that using an average of speckle patterns recorded for different fiber bends during image reconstruction can decrease the influence of changing the fiber bend, all the aforementioned methods are not robust to bending and consider a single, static, fiber configuration. It should be noted at this stage that although in principle the light propagation in MMFs is often invertible, raster scanning leads to better results due to better signal-to-noise (SNR) ratios.

To overcome the fiber bending problem and allow the use of MMFs as flexible imaging devices, extensive research has been carried out over the last decade. Caravaca \textit{et al.}~\cite{caravaca2013} proposed a real-time TM measurement technique that allows for light refocusing at very high frame rates by placing a photodetector at the distal end of the fiber. In another example, Farahi \textit{et al.}~\cite{farahi2013} placed a virtual coherent point light source (a beacon) at the distal end to allow for bending compensation while focusing light through MMFs by digital phase conjugation (DPC). Similarly, Gu \textit{et al.}~\cite{gu2015} attached a partial reflector to the distal end of the MMF, which reflects light back to the proximal end, allowing the MMF to be re-calibrated each time the fiber is bent. 
Leveraging optical memory effects, it was claimed that using a guide-star on the distal facet of an MMF, which reports its local field intensity to the proximal facet, along with the estimation of a basis where the TM is diagonal, enables TM approximation and allows for imaging through a bending MMF~\cite{li2021memory}. All these methods require a priori computation of the TM or feedback mechanism from the distal end of the fiber. In another work~\cite{ploschner2015}, it was shown that bending deformations in step-index MMFs could be predicted and compensated for in imaging applications. However, this process is computationally intensive and requires precise knowledge of the fiber layout.

Recently, deep learning approaches~\cite{moran2018,borhani2018,rahmani2018,turpin2018,kakkava2019,caramazza2019,fan2019,li2020,zhao2021,liu2022,mitton2022l} have emerged that infer images from speckle patterns without prior knowledge of the fiber characteristics. Although they demonstrate promising results and real-time reconstruction, accounting for fiber bending remains a challenging problem. In this context, Li \textit{et al.}~\cite{li2018,li2021displacement} showed that training an autoencoder (AE) neural network on speckles from multiple thin diffusers can be used to recover images from speckles corresponding to a new diffuser of the same type. Moreover, another study~\cite{starshynov2022} confirmed the existence of statistical dependencies in the optical field distribution scattered by a random medium. Motivated by these findings, Resisi \textit{et al.}~\cite{resisi2021} extended the work of Li \textit{et al.}~\cite{li2018,li2021displacement} to MMFs and showed that an AE trained on hundreds of fiber bends can be used to reconstruct images from new configurations. However, their results were limited to the reconstruction of handwritten digits, which share very similar features, displayed on a binary DMD. Furthermore, the reflected light from the fiber was imaged on a camera which does not depict a realistic imaging scenario. Finally, preparing the data for training and testing required 14 weeks of lab work.

In this paper, we propose a high-resolution and minimally invasive imaging system leveraging MMFs. In contrast to previously reported work, our proposed framework represents a realistic imaging scenario where no access to the distal end of the fiber is required during imaging and the collected signal need not be coherent light. The object of interest is probed with speckle patterns through an illumination MMF while a secondary collection fiber, placed alongside the illumination fiber, transmits the diffuse reflections from the object back to the proximal end to be recorded by an avalanche photodiode (APD). The object is then reconstructed from the APD measurements leveraging a variational autoencoder with a Gaussian mixture latent space (GMVAE). In contrast to deterministic deep learning approaches proposed in the literature, the GMVAE learns the overall distribution representing the complicated mapping between measurements corresponding to different fiber configurations and the underlying images of different classes in a low-dimensional latent representation. The latent space learned is configuration-agnostic. This means all measurement vectors corresponding to the same class of images but to different fiber configurations share the same cluster in the latent space. Our GMVAE architecture is significantly simpler than the widely used AEs proposed for imaging through MMFs~\cite{resisi2021} and scattering media~\cite{li2018,li2021displacement}, resulting in a 90\% reduction in training time for the same data set. Our light network does not suffer as much from generalization issues as reported for the AE architecture proposed for imaging through MMFs~\cite{resisi2021}. We demonstrate the robustness of our GMVAE against AE~\cite{li2018,li2021displacement} on new image classes from the fashion-MNIST data set and new configurations of a MMF that were not used during training. This research lays the foundation for a flexible, affordable, and high-resolution imaging system, particularly useful in cases highly sensitive to implant size.

\section*{Results}
\label{sec:results}
\begin{figure}[h]%
\centering
\includegraphics[width=0.9\textwidth]{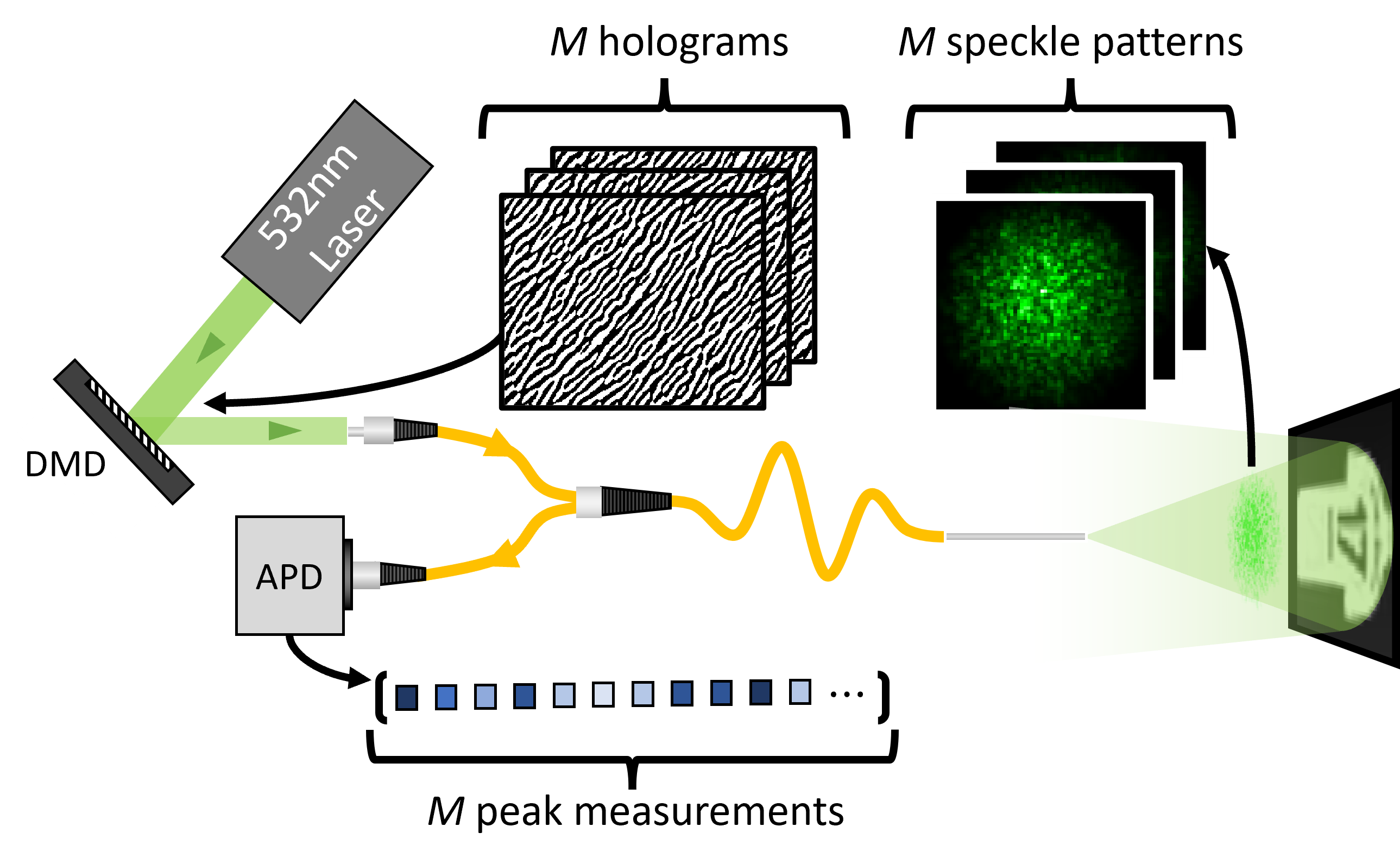}
\caption{\textbf{Simplified experimental setup for reflective imaging through an optical fiber.} Illumination and collection fibers are signified by propagation arrows. DMD - digital micro-mirror device, APD - avalanche photodiode. }\label{fig:setup_apd}
\end{figure}

\subsection*{Observation Model}
\label{sec:results:model}
A simplified version of the proposed imaging system is shown in Figure~\ref{fig:setup_apd}. The laser is spatially shaped by a digital micro-mirror device (DMD) to produce $M$ speckle patterns at the far-field of the distal end of a 300 mm long and 62.5~$\mu$m core graded-index illumination fiber. The $M$ patterns are used to probe a $10~\times~10$~cm object placed approximately at 20~cm from the fiber and the back-scattered light from each of the $M$ patterns is collected by a 400~$\mu$m core step-index collection fiber (running parallel to the illumination fiber) and recorded by an APD at the proximal end (see Methods for details). In contrast to previously reported work, our imaging system does not require access to- or a feedback signal from the distal end of the fiber during imaging.

Assuming that the speckle patterns projected at the distal end of the fiber are known (we will discuss in the training procedure section how they are recorded before actual imaging), each measurement received by the APD can be modeled as the overlap integral between the object and one speckle pattern, hence the forward model can be formulated as
\begin{equation}
\bs{y} = \mathcal{G} (\mathbf{A} \bs{x}) + \bs{n},
\label{eq:forward-model}
\end{equation}
%
where $\bs{y} \in \mathbb{R}^M$ is the measurement vector,  $\mathbf{A} \in \mathbb{R}^{M \times N}$ is the measurement matrix which contains the set of $M$ speckle patterns, vectorized to form the rows of the matrix, each of size $N$ pixels. The vector $\bs{x} \in \mathbb{R}^N$ is a discretized version of the object to be reconstructed and $\bs{n} \in \mathbb{R}^M$ represents measurement noise, modelled as a realization of a random i.i.d. Gaussian noise. The function $\mathcal{G}$ comprises the damping effect caused by the collection fiber and the perturbations of the APD and the laser (see Methods for details). As discussed in the introduction, estimating the underlying object $\bs{x}$ from the measurements $\bs{y}$ yields an inverse problem that can be solved efficiently using a variety of different methods, provided that $\mathbf{A}$ and $\mathcal{G}(\cdot)$ do not change. However, changing the configuration of the fiber means significant changes in the measurement matrix $\mathbf{A}$, and hence the measurement values. Thus, system re-calibration is usually performed to correct for changes in the measurement matrix $\mathbf{A}$. In this work, we aim to learn an $\mathbf{A}$-agnostic variational autoencoder that can classify and reconstruct images from measurements corresponding to new and unseen configurations of the fiber.

\subsection*{Reconstruction algorithm}
\label{sec:results:algo}
\begin{figure}[h!]%
\centering
\includegraphics[width=1\textwidth]{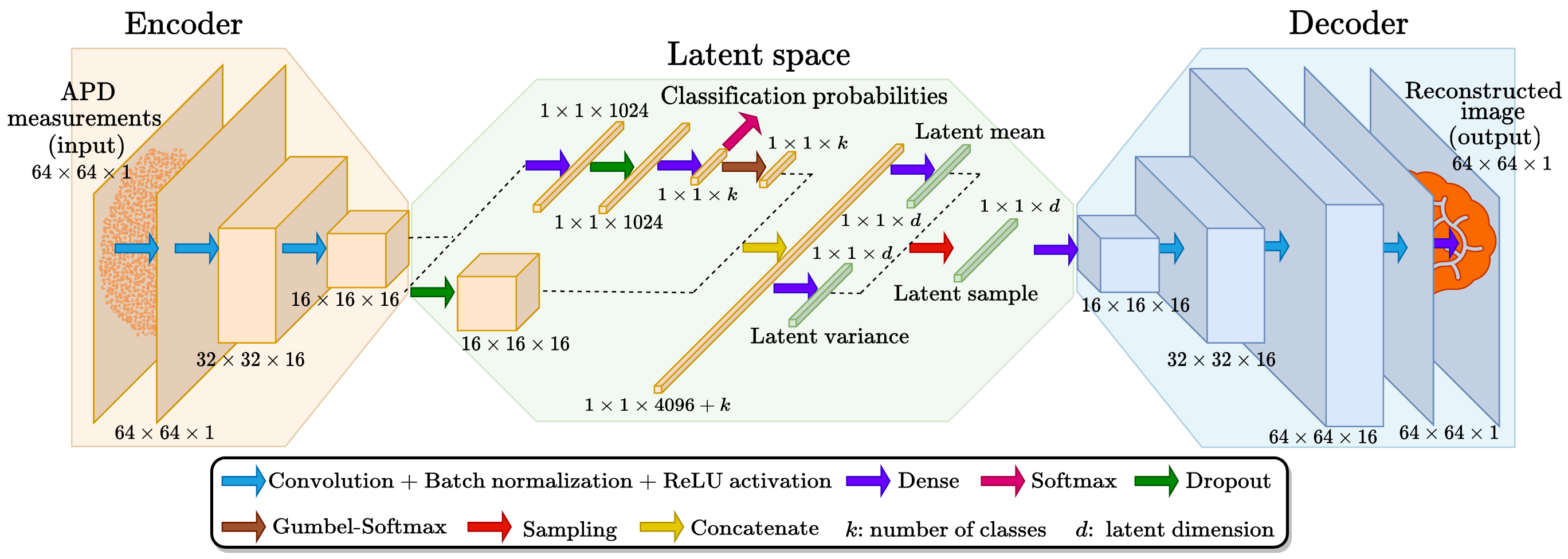}
\vspace*{.1cm}
\caption{\textbf{Schematic of the proposed Gaussian Mixture Variational Autoencoder (GMVAE).} Arrows of different colors indicate various types of neural network layers, as explained by the color-coding in the legend above. The output dimensions for each layer are displayed within the GMVAE schematic. Note that the APD measurement vectors of size $M=4096$ are reshaped to $64\times 64$.}
\label{fig:gmvae}
\end{figure}
Variational autoencoders (VAEs)~\cite{kingma2014, kingma2019, tonolini2020} are deep generative models that consist of an encoder, a latent space and a decoder. The essence of VAEs is to regularize the encodings distribution during training to ensure that the latent space captures only the important features of the data set. This allows for reconstruction (or generation) of new data through the decoder. In contrast to the standard VAE architecture which only contains a single continuous (and multivariate) latent variable, the Gaussian mixture VAE (GMVAE)~\cite{shu2016, figueroa2019,  collier2019,  charakorn2020, varolgunecs2020} also contains a discrete latent variable representing the data class. Therefore, by specifying different integer numbers to different object classes, the GMVAE provides the membership probabilities of the observed object, for all the pre-defined classes, which allows for object classification. 

In this context, we leverage a GMVAE (depicted in Figure~\ref{fig:gmvae}) for imaging through a MMF which bends. Essentially, the encoder of our GMVAE performs non-linear dimensionality reduction of a set of APD measurements corresponding to any image and any fiber configuration into a low-dimensional latent representation which only captures image information and disregards the fiber configuration. Thus, the learned latent space is trained to be $\mathbf{A}$-agnostic, i.e., all measurement vectors corresponding to the same image but different fiber configurations share the same features in the latent space. The decoder of the GMVAE then takes as input this latent vector and generates an estimated image of the scene. Once trained, the proposed GMVAE can simultaneously classify (through the VAE encoder) and reconstruct (through the VAE decoder) objects from measurement vectors corresponding to a variety of fiber configurations, even configurations unseen during training. 

\subsection*{Training procedure}\label{sec:results:training}
\begin{figure}[h!]
\centering
\includegraphics[width=0.9\textwidth]{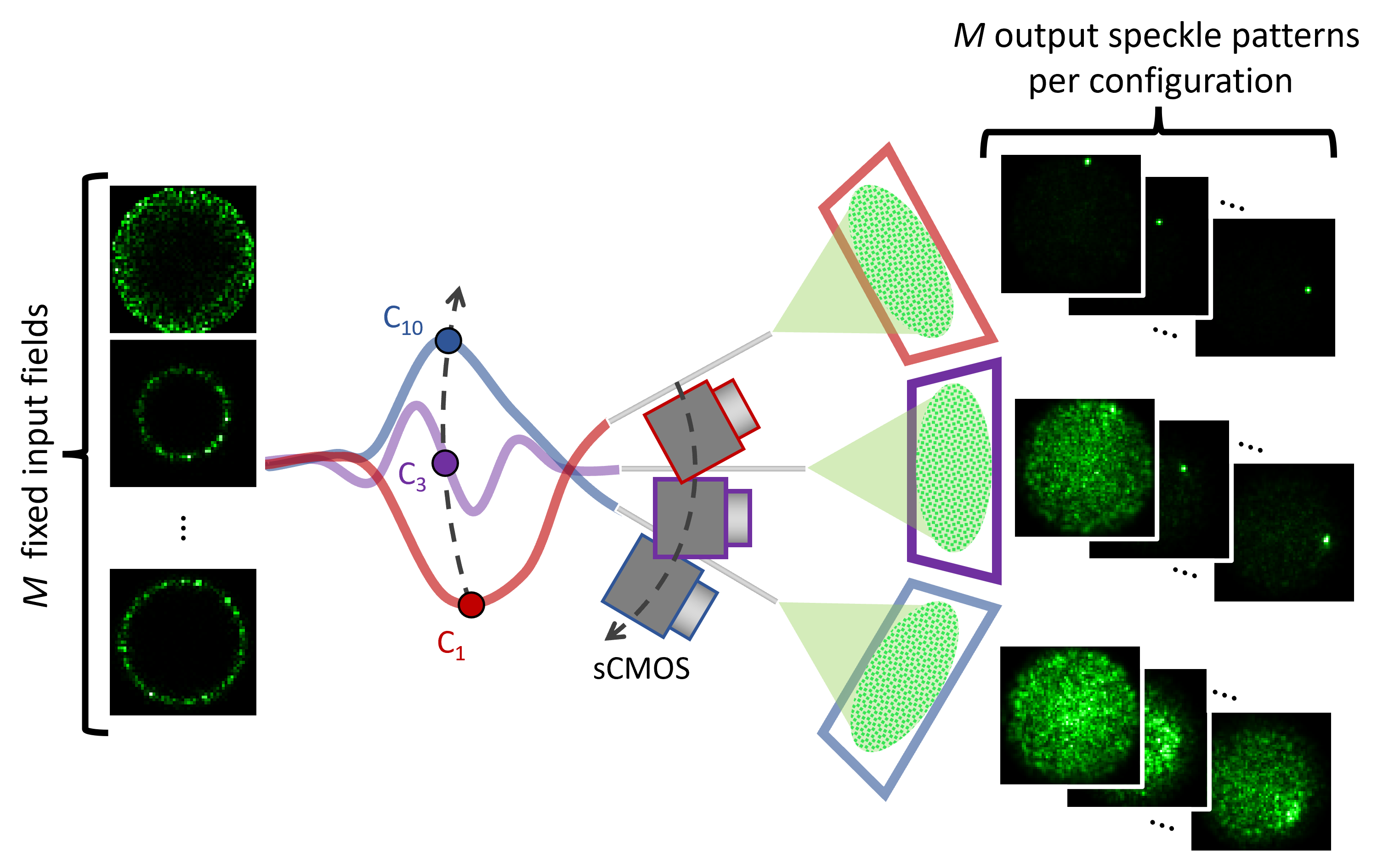}
\caption{\textbf{Optical setup for collecting the training data set.} The figure shows three simplified fiber configuration examples and the resulting changes to the output speckle patterns. The fiber tip, imaging screen, and sCMOS camera were rotated together using the rotation stage and the middle of the fiber was clamped to- and moved by the bending arm (see Supplementary Note 1 for details). }
\label{fig:setup_training}
\end{figure}

Recall that training the GMVAE described above requires sets of input (APD measurements) and output (reference images of objects, belonging to different classes). However, recording sufficient measurement vectors at different fiber configurations and with a large image data base for training is not possible in practice as this imaging system is based on a sequential generation of illumination patterns in order to record diffuse reflection from real objects. To circumvent this problem, we use a white screen and a scientific CMOS camera (Hamamatsu Orca Flash 4.2) (sCMOS) to record the speckle patterns corresponding to $M$ fixed DMD patterns for different configurations of the fiber. This leads to different speckle matrices $\mathbf{A}_l$, where each speckle is a row vector in $\mathbf{A}_l$ and $l$ represents the index of the fiber configuration. As illustrated in Figure~\ref{fig:setup_training}, the DMD patterns, and therefore input spatial light field, are maintained for all configurations which allows the model to capture the hidden correlations between speckle patterns corresponding to the different configurations. We then use the recorded $\mathbf{A}_l$ matrices to perform speckle projection on images numerically to simulate measurement vectors. More precisely, we compute $\bs{y}$ as in Equation~\eqref{eq:forward-model} for each image $\bs{x}$ in our training data set and for the $L$ configurations of the fiber (replacing $\mathbf{A}$ in Equation~\eqref{eq:forward-model} by $\mathbf{A}_l$).

To assess quantitatively the performance of the model in the next section, we first generate data sets in a similar fashion to the training data, however, we record speckle patterns corresponding to new configurations of the fiber which are different from those seen during training. We then perform projection on the test images numerically to simulate measurements. Measurements computed with this procedure are referred to as numerical measurements. We also assess qualitatively the performance of GMVAE on real measurements collected by an APD as presented in Figure~\ref{fig:setup_apd}. We refer to these measurements as APD measurements. Training using numerical measurements offers the benefit of greater speed and flexibility to the model as learning new classes of images does not require the lab experiments to be repeated. 

\subsection*{Experiments}\label{sec:results:experiments}
In the following experiments, we apply two types of bend simultaneously on the fiber. These are rotational bends applied at the final section of the fiber using a rotational stage, and arm bends performed on the middle section of the fiber leveraging a large lever arm (the experimental setup is shown in Supplementary Note 1). The lever arm has a mechanical advantage of $\sim$1:8 from the actuation side to the fiber-clamp side. Arm bends were measured at the actuation side in -1 mm increments from the initial position at 10 mm to 0 mm while rotational bends were in 5$^\circ$ increments from the initial position at 230$^\circ$ to 280$^\circ$. This yielded $L=11$ different fiber configurations covering a change in fiber bend of 50$^\circ$ and range of movement of 8 cm. For simplicity, we denote each configuration by $\text{C}_x$ where $x$ is the corresponding arm bend position.

GMVAE training was performed using $L=5$ fiber configurations ($\text{C}_{10}$, $\text{C}_{7}$, $\text{C}_{5}$, $\text{C}_{3}$ and $\text{C}_{1}$). These configurations cover the range from (10~mm, 230$^\circ$) to (1~mm, 275$^\circ$). At each configuration, we use 6000 images from each of the following 8 classes of the fashion-MNIST data set: (0) t-shirt, (1) trouser, (2) dress, (3) coat, (4) sandal, (5) shirt, (6) sneaker, and (7) boot. Testing was done on new configurations ($\text{C}_{9}$, $\text{C}_{8}$, $\text{C}_{6}$, $\text{C}_{4}$, $\text{C}_{2}$, $\text{C}_{0}$) using new images from the same trained-on 8 classes and 2 new classes: pullover and bag. Here, $\text{C}_{9}$, $\text{C}_{8}$, $\text{C}_{6}$, $\text{C}_{4}$, $\text{C}_{2}$ represent new bends which lie inside the range of the training configurations while $\text{C}_{0}$ (0~mm,  280$^\circ$) represents a new bend which lies outside the range of the training configurations.

We tested the method in two scenarios, with and without wavefront shaping. Note that wavefront shaping allows for raster-scanning imaging while no wavefront shaping leads to speckle imaging. The performance of GMVAE is evaluated in terms of reconstruction quality against AE~\cite{li2018,li2021displacement} on numerical and APD measurements. Note that numerical measurements are computed at all $L=11 $ configurations while the APD measurements are only recorded at $\text{C}_{10}$, $\text{C}_{7}$, $\text{C}_{5}$, $\text{C}_{2}$ and $\text{C}_{0}$. Since AE can only perform reconstruction, the classification accuracy of GMVAE is compared to that of a classifier trained on the images reconstructed by AE. For fairness, the chosen classifier, denoted by C-AE, has the same architecture as that used in our GMVAE (see Supplementary Note 4 for details).

\subsubsection*{First experiment with wavefront shaping}\label{sec:results:first-experiment}
In this experiment, we use the wavefront shaping technique~\cite{stellinga2021} to generate focal spots at the distal end of the fiber. This allows for raster-scanning imaging which can give better results due to better SNR ratios. Speckles were recorded at the different configurations and wavefront shaping was only performed at the configuration $\text{C}_{10}$. This means that speckles at $\text{C}_{10}$ (10~mm,  230$^\circ$) are simple focal points. By bending the fiber away from the calibration position, the focal points become speckles due to coupling between spatial modes of the fiber. Therefore, $\text{C}_{0}$  (0~mm,  280$^\circ$) represents the most challenging bend where speckles have the maximum distortion (see Supplementary Figure 2).

Figure~\ref{fig:psnr_curve}, \textbf{(A)} shows the peak signal-to-noise ratio (PSNR) curves obtained by the GMVAE and the AE at all configurations under scrutiny utilizing 10000 numerical measurements and 20 APD measurements from the fashion-MNIST dataset (8 trained-on classes and 2 new classes). We notice that both the GMVAE and the AE give the best reconstruction quality on the numerical measurements recorded at $\text{C}_{10}$ (the AE scores PSNR = $23.67\pm3.79$ dB and the GMVAE scores PSNR = $22.69\pm3.61$ dB). This is expected since $\text{C}_{10}$ was used during training. Moreover, $\text{C}_{10}$ is the calibrated configuration, hence details from the original images can be seen directly in the measurements, reducing the problem of reconstruction to a denoising or a mild deconvolution problem. Although the AE was able to achieve 1 dB enhancement on numerical measurements at the calibrated configuration, the GMVAE achieves higher PSNR values at all other configurations. In contrast to the AE which exhibits a gradual decrease in PSNR values on numerical measurements as the fiber bends away from the calibrated position, the GMVAE maintains almost the same PSNR values ($\thickapprox 20$ dB) at the seen configurations ($\text{C}_{7}$, $\text{C}_{5}$, $\text{C}_{3}$ and $\text{C}_{1}$) and the same PSNR values ($\thickapprox 18.5$ dB) at the unseen configurations ($\text{C}_{9}$, $\text{C}_{8}$, $\text{C}_{6}$, $\text{C}_{4}$, $\text{C}_{2}$ and $\text{C}_{0}$). Albeit not significant, the difference in PSNR values between seen and unseen configurations suggests that not all the configuration-dependent features are discarded in the latent space, hence training the GMVAE with more configurations might lead to better generalization. The GMVAE also maintains around 1.5 dB of PSNR enhancement on APD measurements recorded at the seen configurations ($\text{C}_{10}$, $\text{C}_{7}$ and $\text{C}_{5}$) and the unseen configurations ($\text{C}_{2}$ and $\text{C}_{0}$). Since both the AE and the GMVAE were trained on numerical measurements, we notice a steep decrease in the PSNR values when testing on APD measurements. Although the training procedure was adapted to partially compensate for the damping effect caused by the collection fiber and the perturbations of the APD and the laser (see Methods for details), small additional variations occur in the APD measurement. On the one side, this suggests the use of a more robust hardware and possibly enhance the forward model to improve the training efficiency. On the other hand, this demonstrates the performance of the methods in realistic scenarios including unknown signal variations.

Figure~\ref{fig:images_raster} compares qualitatively the reconstruction quality of the GMVAE against the AE on APD measurements collected at the following configurations: $\text{C}_{10}$,  $\text{C}_{2}$ and $\text{C}_{0}$. We can clearly see the good reconstruction quality for both the GMVAE and the AE at the calibrated configuration $\text{C}_{10}$. Again, the results show a superior performance for the GMVAE in comparison to the AE for the new configurations ($\text{C}_{2}$ and $\text{C}_{0}$) which are far away from the calibration position. Although $\text{C}_{0}$ represents the most difficult bend which lies outside the range of training configurations, the GMVAE was able to maintain almost the same reconstruction quality as that for $\text{C}_{2}$ which lies inside the training range. This shows the ability of the GMVAE to generalize to new configurations of the fiber in comparison with the AE that fails in capturing the non-linear dependencies between the different configurations. Moreover, the training time of the GMVAE is around 10 times lower than that for the AE.

Figure~\ref{fig:confusion_raster} (top row) shows the average confusion matrices obtained by the GMVAE and the C-AE on 8000 numerical measurements from the 8 trained-on classes. The confusion matrices are averaged over the unseen configurations $\text{C}_{9}$, $\text{C}_{8}$, $\text{C}_{6}$, $\text{C}_{4}$, $\text{C}_{2}$ and $\text{C}_{0}$. We can see that the GMVAE achieves $77\pm3\%$ classification accuracy with $1\%$ average enhancement over the C-AE which scores $76\pm8\%$. These numbers suggest that simultaneous learning is better than sequential learning and can boost both reconstruction quality and classification accuracy for unseen configurations of the fiber.
 
Figure~\ref{fig:pca_raster} (top row) presents the combined raw data and the combined GMVAE latent vectors of 8000 numerical measurements computed at the configurations $\text{C}_{10}$, $\text{C}_{2}$ and $\text{C}_{0}$ and projected using principal component analysis (PCA) to the 3D space. In contrast to the combined raw data (Figure~\ref{fig:pca_raster}, \textbf{(A)}), the GMVAE combined latent space (Figure~\ref{fig:pca_raster}, \textbf{(B)}) of the testing data points shows a clear separation between the different classes. More importantly, the combined latent space is configuration-agnostic. 
This means that regardless of the varying configurations ($\text{C}_{10}$, $\text{C}_{2}$, and $\text{C}_{0}$), the measurement vectors belonging to the same class (e.g., t-shirt) coalesced into a single cluster in the latent space. This observation is a manifestation of the GMVAE's ability to capture the commonalities in the underlying data structure corresponding to the same class, irrespective of the configuration changes.
\begin{figure}[h!]
\centering
\begin{subfigure}{0.49\textwidth}
\centering
\small{\textbf{(A)} with wavefront shaping}\\
\includegraphics[width=1\linewidth]{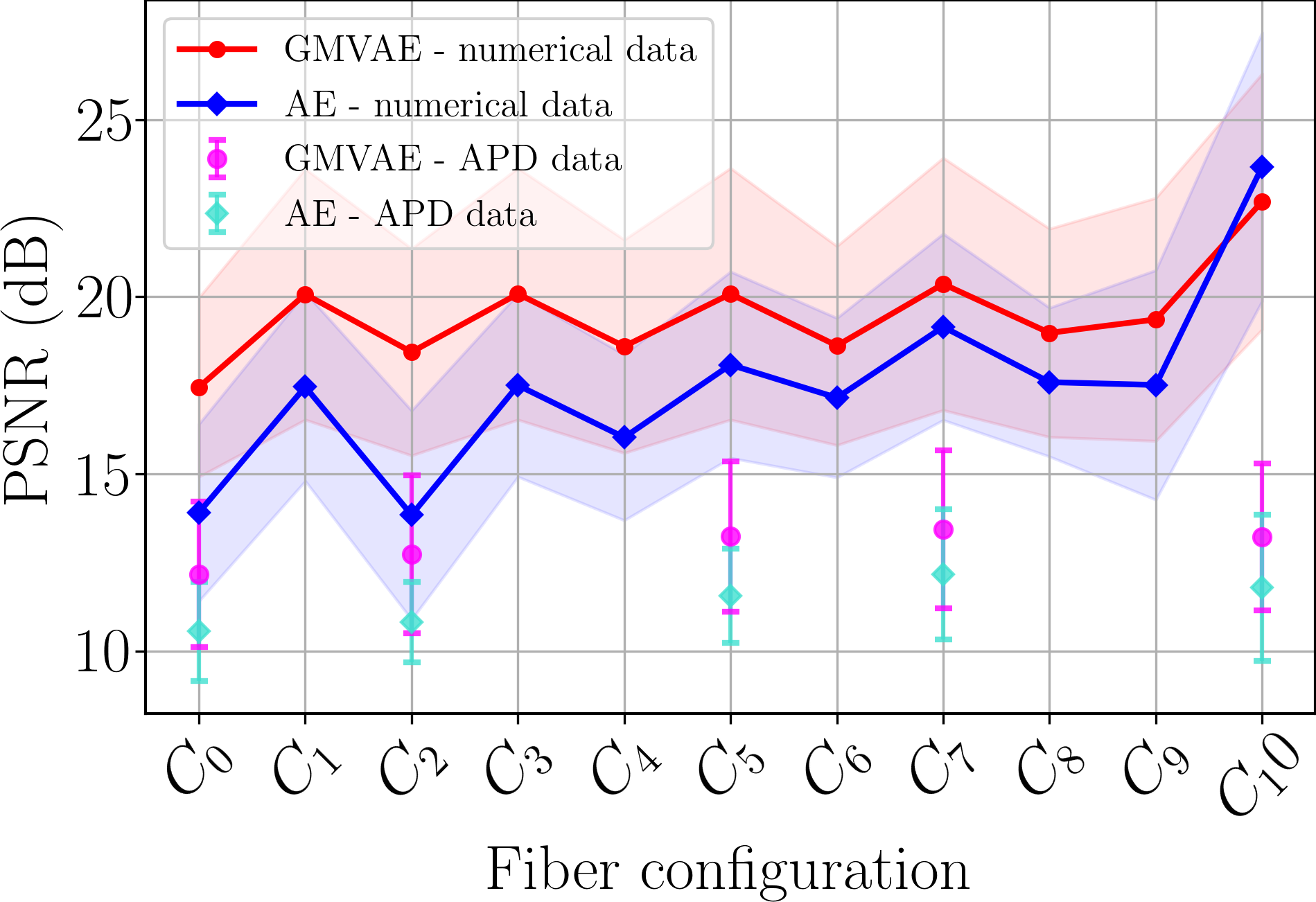}\\
\end{subfigure}%
~
\begin{subfigure}{0.49\textwidth}
\centering
\small{\textbf{(B)} without wavefront shaping}\\
\includegraphics[width=1\linewidth]{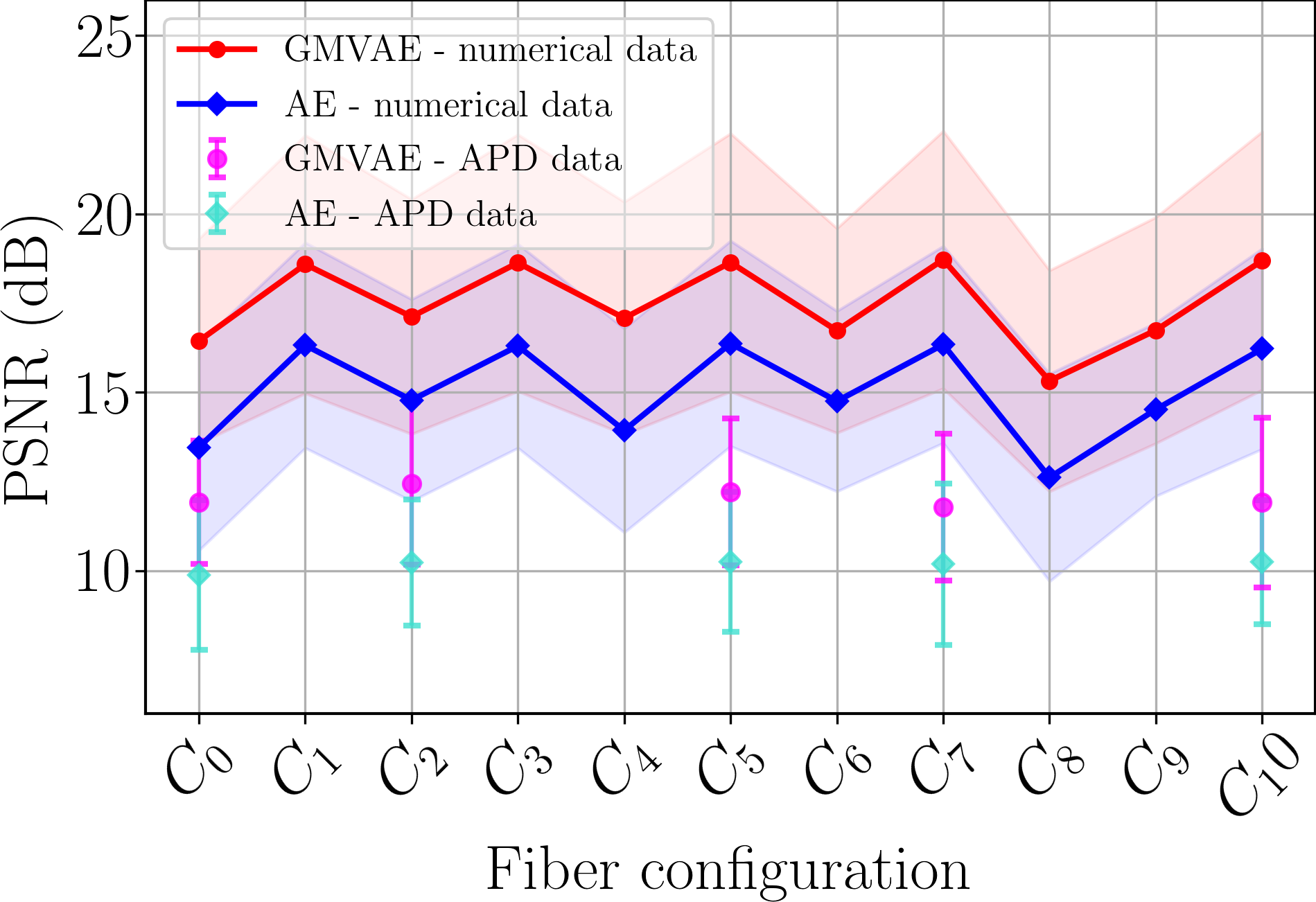}\\
\end{subfigure}%
~
\caption{\textbf{Average PSNR curves obtained by GMVAE and AE.} Results are shown at different configurations on 10000 numerical measurements and 20 APD measurements from the fashion-MNIST dataset (8 trained-on classes and 2 new classes). \textbf{(A)} results of the first experiment with wavefront shaping and \textbf{(B)} results of the second experiment without wavefront shaping.}
\label{fig:psnr_curve}
\end{figure}

\begin{figure}[h!]
\centering
\includegraphics[trim={2.2cm 16cm 2.2cm 1.6cm},clip,width=1\linewidth]{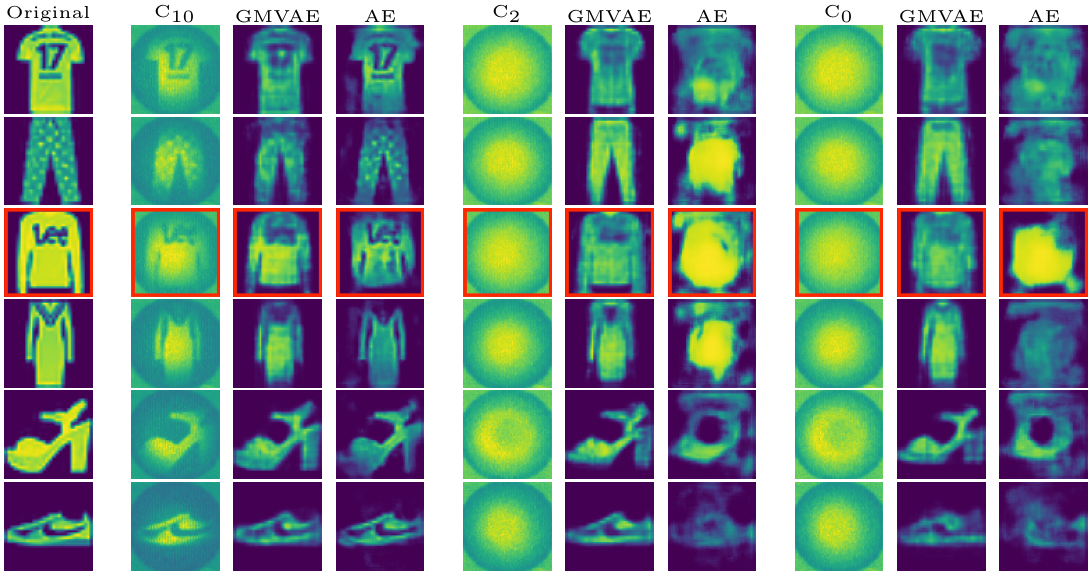}\\
\caption{\textbf{Reconstruction results from the first experiment with wavefront shaping at configurations $\text{C}_{10}$, $\text{C}_{2}$ and $\text{C}_{0}$.} The first column displays the original images, while each set of three consecutive columns presents the APD measurements, GMVAE reconstruction, and AE reconstruction. Images highlighted by red squares correspond to a new class (pullover) not used during training. The intensity values of all images in the figure range between 0 and 1.}
\label{fig:images_raster}
\end{figure}
\begin{figure}[h!]
\centering
\begin{subfigure}{0.48\textwidth}
\centering
\small{\textbf{(A)} 1\textsuperscript{st} experiment, GMVAE ($77\pm 3\%$)}\\
\includegraphics[width=1\linewidth]{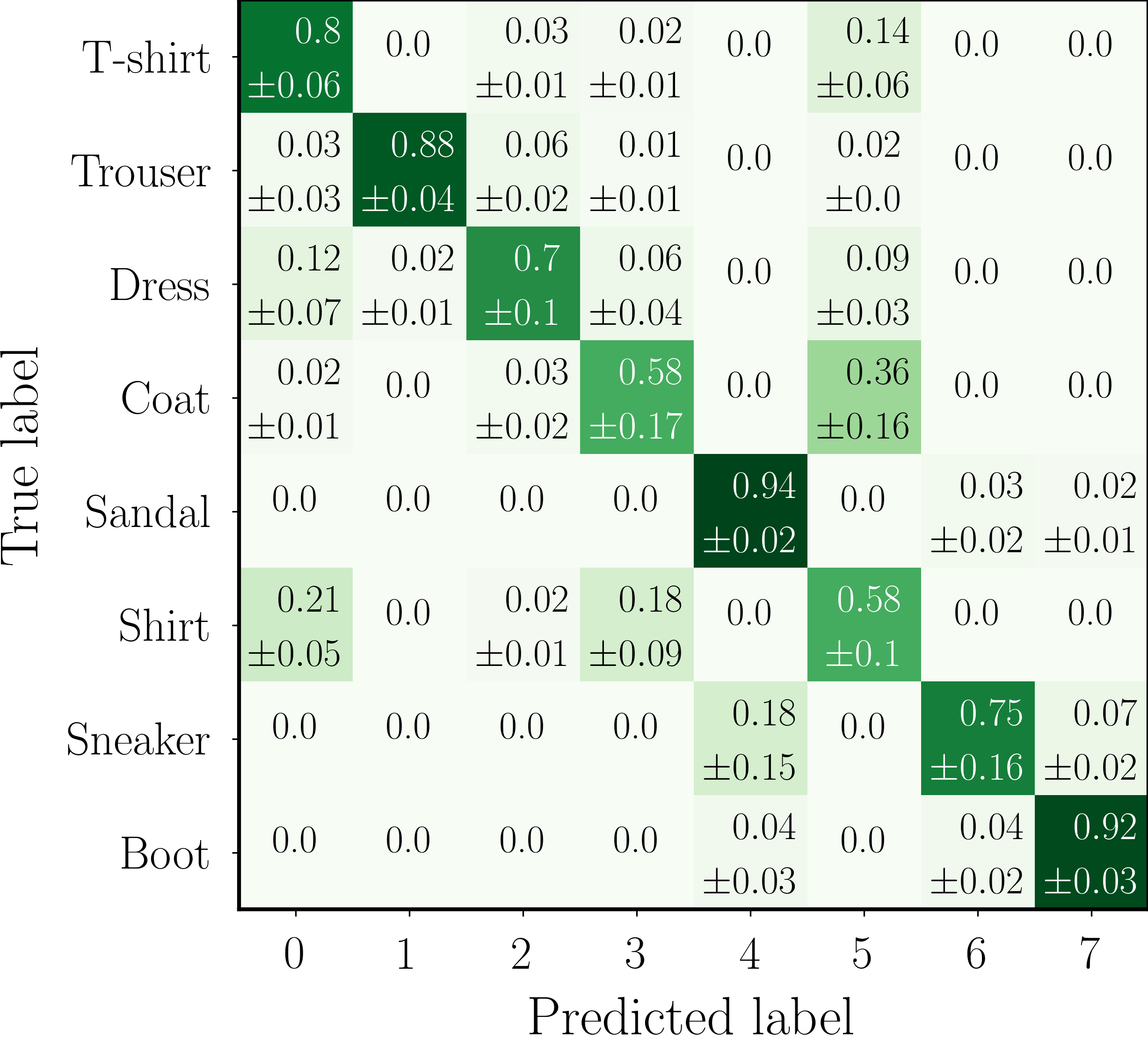}\\
\small{\textbf{(C)} 2\textsuperscript{nd} experiment, GMVAE ($66\pm 10\%$)}\\
\includegraphics[width=1\linewidth]{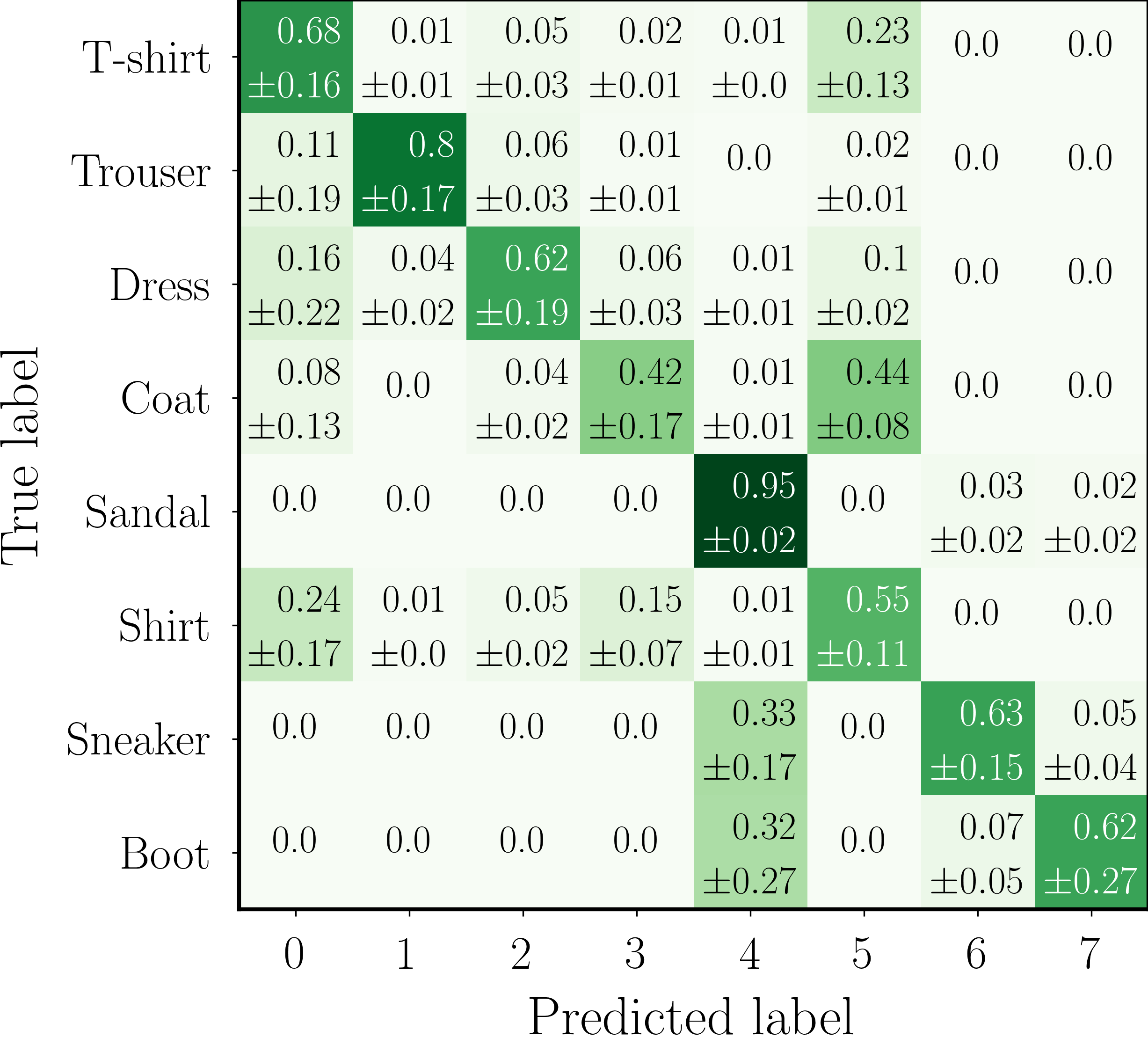}\\
\end{subfigure}%
~
\begin{subfigure}{0.48\textwidth}
\centering
\small{\textbf{(B)} 1\textsuperscript{st} experiment, C-AE ($76\pm 8\%$ )}\\
\includegraphics[width=1\linewidth]{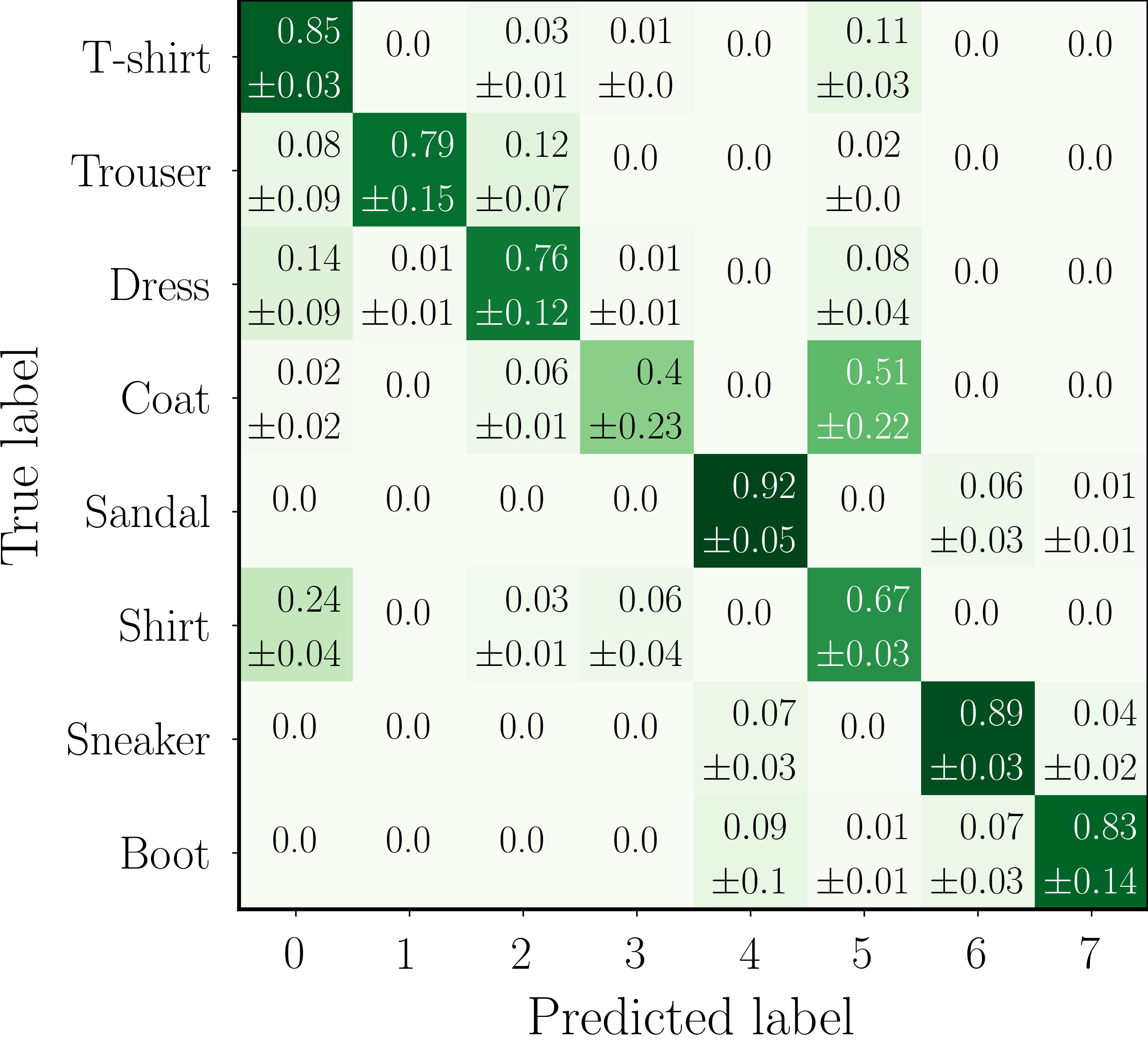}\\
\small{\textbf{(D)} 2\textsuperscript{nd} experiment, C-AE ($58\pm 8\%$)}\\
\includegraphics[width=1\linewidth]{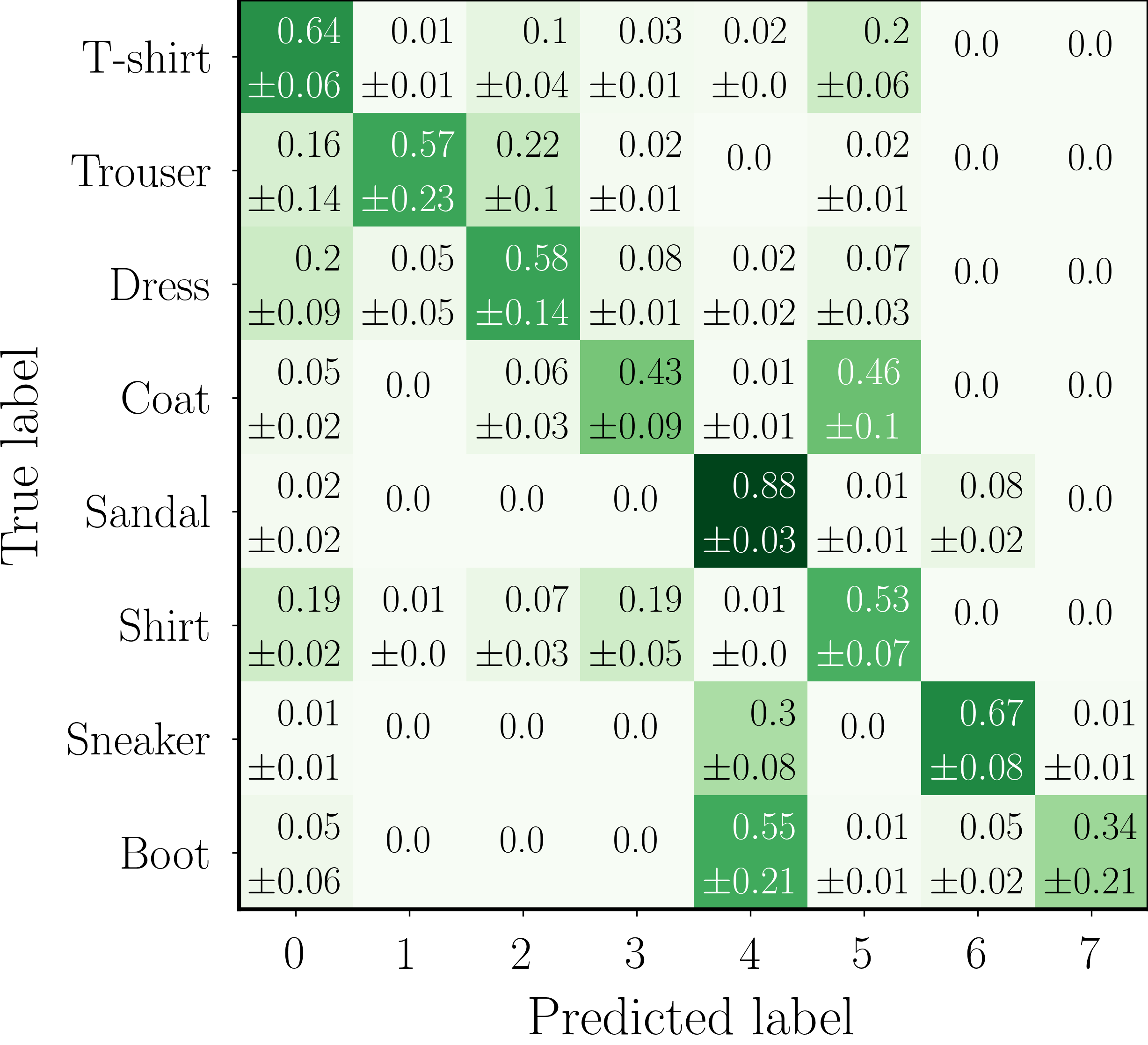}\\
\end{subfigure}%
~
\caption{\textbf{Average confusion matrices obtained by GMVAE and C-AE on 8000 numerical measurements from the 8 trained-on classes.} \textbf{(A)} and \textbf{(B)} are results of the first experiment with wavefront shaping, and \textbf{(C)} and \textbf{(D)} are results of the second experiment without wavefront shaping. The confusion matrices are averaged over the unseen configurations  $\text{C}_{9}$, $\text{C}_{8}$, $\text{C}_{6}$, $\text{C}_{4}$, $\text{C}_{2}$ and $\text{C}_{0}$.}
\label{fig:confusion_raster}
\end{figure}
\begin{figure}[h!]
\centering
\begin{subfigure}{0.48\textwidth}
\centering
\small{\textbf{(A)} 1\textsuperscript{st} experiment, raw data}\\
\includegraphics[width=1\linewidth]{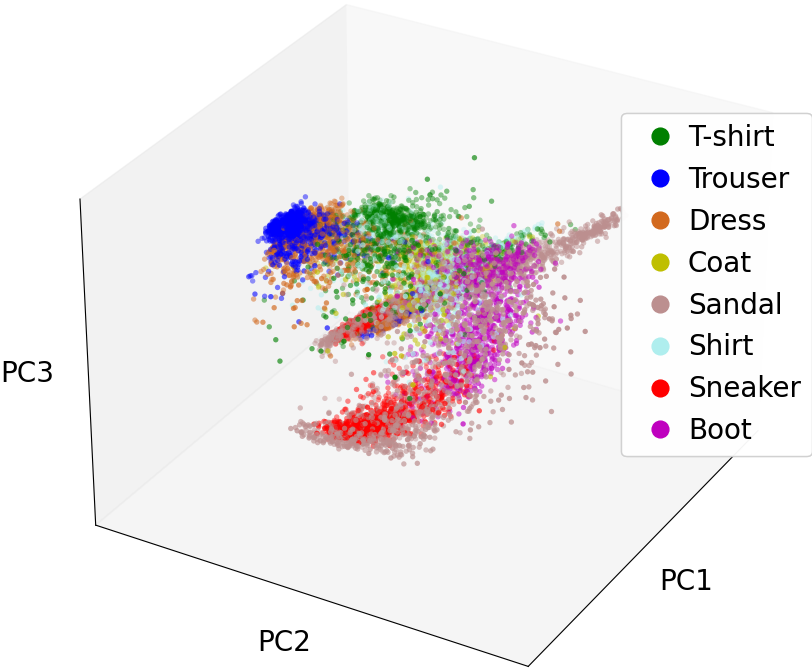}\\
\vspace{0.3cm}
\small{\textbf{(C)} 2\textsuperscript{nd} experiment, raw data}\\
\includegraphics[width=1\linewidth]{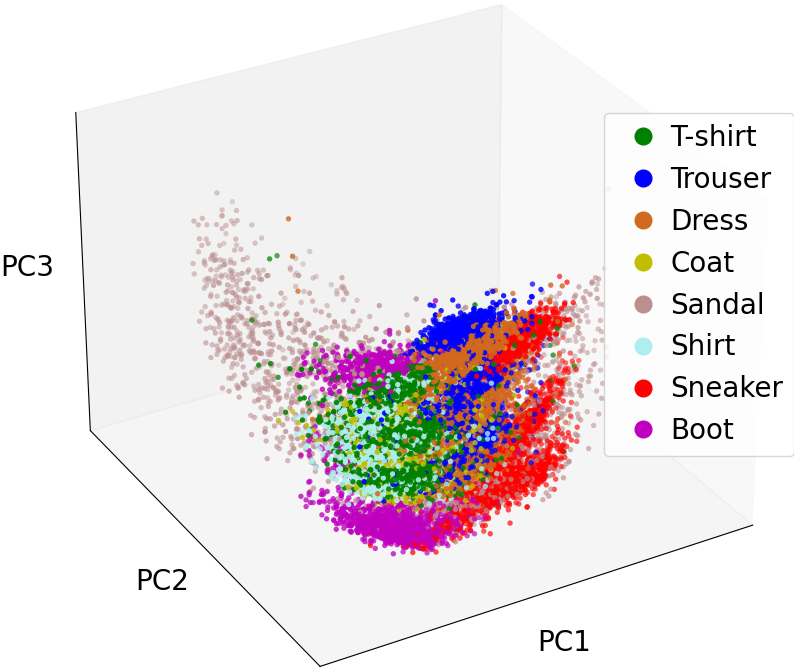}\\
\end{subfigure}%
~
\begin{subfigure}{0.48\textwidth}
\centering
\small{\textbf{(B)} 1\textsuperscript{st} experiment, latent space}\\
\includegraphics[width=1\linewidth]{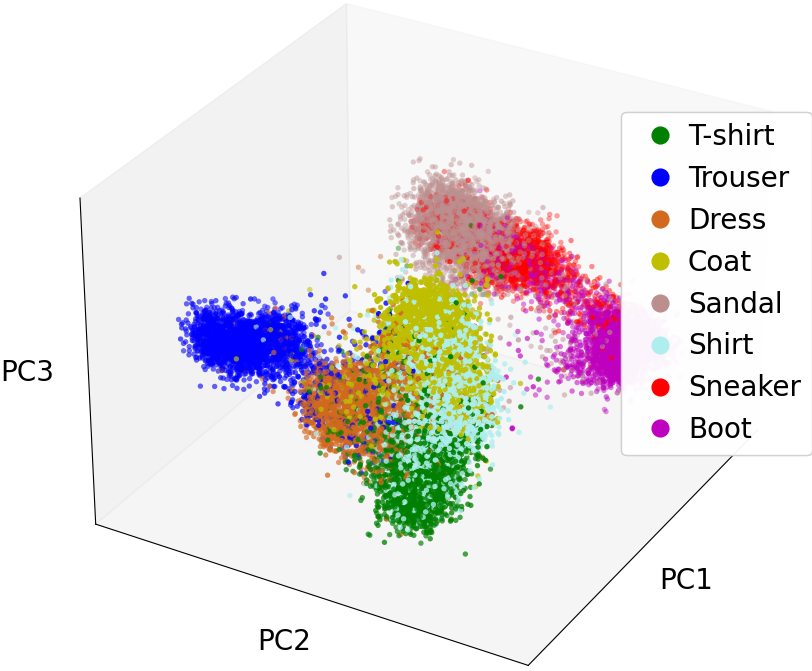}\\
\vspace{0.3cm}
\small{\textbf{(D)} 2\textsuperscript{nd} experiment, latent space}\\
\includegraphics[width=1\linewidth]{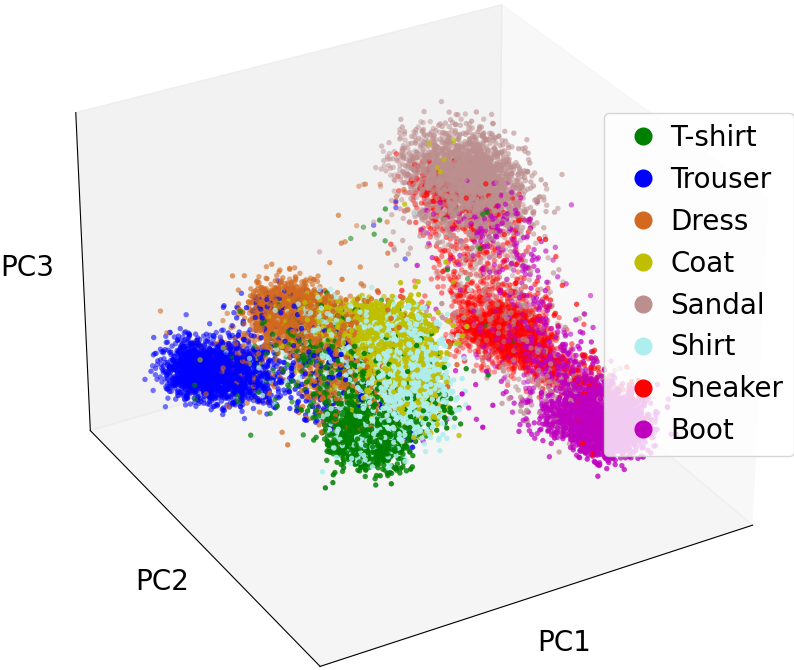}\\
\end{subfigure}%
~
\caption{\textbf{3D PCA projection of the raw data and the GMVAE latent vectors.} \textbf{(A)} and \textbf{(B)} are results of the first experiment with wavefront shaping, and \textbf{(C)} and \textbf{(D)} are results of the second experiment without wavefront shaping. Each panel shows the accumulation of 8000 numerical measurements computed at the configurations $\text{C}_{10}$, $\text{C}_{2}$ and $\text{C}_{0}$.}
\label{fig:pca_raster}
\end{figure}

\subsubsection*{Second experiment without wavefront shaping}\label{sec:results:second-experiment}
In the second experiment, speckles were recorded at the different configurations with no configuration-specific wavefront shaping performed. This simulates a situation in which no prior calibration has been done and the fiber is to be used in-situ for imaging directly. This experiment represents a more complicated scenario as all configurations are challenging (see Supplementary Figure 2). However, it could be of interest when wavefront shaping for raster scanning is expensive or not possible.

Figure~\ref{fig:psnr_curve}, \textbf{(B)} presents the PSNR curves obtained by the GMVAE and the AE at all configurations on numerical and APD measurements. We see that the GMVAE achieves higher PSNR values (around 2 dB) for all configurations on both numerical and APD measurements. The GMVAE gives almost the same PSNR values at the seen configurations and the same PSNR values at the unseen configurations (except for $\text{C}_{8}$). The drop in PSNR values at the unseen configurations indicates that the GMVAE latent space does not only capture image information but also fiber configuration related information. This problem might be mitigated by increasing the number of training configurations.

The average confusion matrices obtained by the GMVAE and the C-AE on 8000 numerical measurements from the 8 trained-on classes are shown in Figure~\ref{fig:confusion_raster} (bottom row). The confusion matrices are averaged over the unseen configurations. Once again, the results suggest better classification accuracy for the GMVAE compared to the C-AE. More precisely, the GMVAE scores $66\pm10\%$ with 8\% average enhancement from the C-AE.

Finally, the combined raw data and the combined GMVAE latent vectors of 8000 numerical measurements computed at the configurations $\text{C}_{10}$, $\text{C}_{2}$ and $\text{C}_{0}$ and projected by PCA to the 3D space are presented in Figure~\ref{fig:pca_raster} (bottom row). In contrast to the combined raw data (Figure~\ref{fig:pca_raster}, \textbf{(C)}) where we can see stripes representing measurements from the same class but different configurations, the combined GMVAE latent space (Figure~\ref{fig:pca_raster}, \textbf{(D)}) is configuration-agnostic, i.e., only features of the measurements related to the underlying images are preserved and configuration-dependent features are discarded.

\subsubsection*{Video reconstruction}\label{sec:results:video}
We further validate the performance of the GMVAE on recorded videos in two scenarios. Firstly, we show the reconstruction of 3 static objects while moving the fiber from the calibrated position $\text{C}_{10}$ (10~mm, 230$^\circ$) to $\text{C}_{5}$ (5~mm, 255$^\circ$). This corresponds to change in fiber bend of 25$^\circ$ and range of movement of 4 cm. Secondly, we show the reconstruction of a moving object while bending the fiber from $\text{C}_{10}$ to $\text{C}_{5}$. The reconstructed video (provided as a supplementary material) shows good reconstruction quality for the GMVAE. As seen in the reconstructed video, it is expected that the GMVAE fails to reconstruct images when objects leave the field-of-view (FOV) of the fiber. We believe this could be improved using more complex VAE architectures allowing translation-blind classification~\cite{franceschi2020, wu2021, wang2022}.

Our proposed system facilitates the reconstruction of roughly 5 frames per second. Despite our GMVAE enabling swift inference times---typically in the millisecond range once properly trained---our setup's real-time potential is predominantly constrained by the hardware and the acquisition process. Our system uses a DMD that reads binary masks from onboard DDR RAM at a rate of 21000 masks per second. To the best of our knowledge, this represents the current state of the art in arbitrary digital projection. With an FPGA upgrade and a restricted region-of-interest on the DMD, we could potentially increase this rate to roughly 40000 masks per second. Given that we scan through 4096 DMD patterns per image, we can achieve an image acquisition rate just above 5 Hz. We are actively investigating potential optimizations for the hardware, firmware, and software used in our system.
It's important to clarify that while our system can perform live reconstruction, we did not showcase this capability in this work, as the processing was conducted offline.

\section*{Discussion}
We have proposed an innovative real-time imaging system using flexible MMFs, which, in contrast to previous work, is robust to bending and does not require access to the distal end of the fiber during imaging. We utilized a MMF 300~mm long with a 62.5~$\mu$m core for imaging $10~\times~10$~cm objects placed approximately at 20 cm from the fiber and the system could deal with a change in fiber bend of 50$^\circ$ and range of movement of 8~cm. Objects from different classes were efficiently classified and reconstructed from the speckle measurements leveraging a Gaussian mixture variational autoencoder (GMVAE). The essence of our GMVAE is to learn a configuration-agnostic latent representation from measurements corresponding to different configurations of the fiber. The complexity of our method is much smaller than that of other works proposed for imaging through MMFs~\cite{resisi2021} and scattering media~\cite{li2018,li2021displacement} and as a result the training time is reduced by around 90\% for the same training data set. The results demonstrated through different experiments with and without wavefront shaping proved the efficiency of the proposed approach in reconstruction and classification of images for new configurations of the fiber. The good performance of our system is further validated on recorded videos of static and moving objects while dynamically bending the fiber, enabling reconstruction of approximately 5 frames per second. This novel approach paves the way for real-time, flexible and high-resolution imaging system for use in areas with very limited access.

\section*{Methods}
\label{sec:methods}

\subsection*{Experimental design}
\label{sec:methods:setup}
The experimental setup, shown in Figure~\ref{fig:setup_apd}, consists of a Q-switched laser at 532 nm with a pulse width of 700 ps and a repetition rate of 21 kHz (Teem Photonics SNG-100P-1x0). The laser is spatially shaped by a digital micro-mirror device (Vialux V-7000) (DMD) such that the far-field of the distal end of a 62.5~$\mu$m core graded index fiber with a numerical aperture (NA) of 0.275 (Thorlabs GIF625) has a desired spatial intensity pattern. 
Of particular note is the fact that the transmission matrix of a graded index fiber is more resistant to bending than that of a step index fiber. Given that our study aims to maximize the bend resistance of our imaging method, we have chosen to use a graded index fiber as our illumination fiber, in conjunction with a bend-agnostic reconstruction approach.
Through calibration of the fiber in a particular configuration, this spatial pattern was selected to be a raster scanning spot as is described in the first experiment with wavefront shaping. As discussed in the second experiment without wavefront shaping, we also investigated the case where the holograms are not selected for any specific configuration of the fiber and, hence, the DMD masks were generated with a quasi-random algorithm. For imaging, a sample image, printed in grey-scale on white paper, is placed at the screen and the reflected light for each of $M$ patterns generated by the DMD is collected by a second step index collection fiber, 400~$\mu$m core and 0.39 NA (Thorlabs FT400UMT), running parallel to the illumination fiber. The back-scattered light is recorded at the proximal end of the collection fiber by an AC-coupled avalanche photodiode (MenloSystems APD210) (APD) sampling at 2.5 Gs/s. The laser pulses are used a trigger for imaging and as such allow for time gating the signal from the collection fiber. In this way, only data corresponding to the distance of the sample from the fiber tip are considered and the remaining recording could be dismissed, hence reducing background illumination effects. After this temporal gating is performed, $M$ speckle measurements remain for each image.

\subsection*{GMVAE training}
\label{sec:methods:training}
The GMVAE architecture is explained in Supplementary Note 3. For both experiments, the training data set utilizes 48000 images of size $64\times 64$ pixels from 8 classes of the fashion-MNIST data set (6000 images per class). We record the speckle patterns corresponding to $M=4096$ random but fixed DMD patterns for $L=11$ configurations of the fiber leading to different speckle matrices $\mathbf{A}_l$, where each speckle pattern of spatial resolution $64\times 64$ is stretched as a row vector in $\mathbf{A}_l$ and $l$ represents the fiber configuration. Then, we perform projections on the training images numerically to get the measurement vectors $\bs{y}$ as in Equation~\eqref{eq:forward-model}.

By expressing the function $\mathcal{G}$ in a matrix form, Equation~\eqref{eq:forward-model} can be re-written as
%
\begin{equation}
\bs{y} =
\begin{bmatrix}
\overline{\mathbf{A}\bs{x}},\quad
~\overline{\left(\frac{\mathbf{A}\bs{x} / s - ~\bs{b}} {\bs{w}-\bs{b}}\right)}
\end{bmatrix}^T
+ \bs{n},
\label{eq:forward-model-2}
\end{equation}
%
where $s$ accounts for the damping effect caused by the collection fiber and is fixed to 10 for the first experiment and 200 for the second experiment, and $T$ represents the transpose operator. 
The vectors $\bs{w}$ and $\bs{b}$ are the white and black backgrounds recorded during the experiments.
Finally,  $\overline{\mathbf{A}\bs{x}}$ is a normalized version of $\mathbf{A}\bs{x}$ with values in the range [0,1].
We found that incorporating these two types of normalization in the training can mitigate the effects of the the perturbations of the APD and the laser and enhance both the reconstruction quality and the classification accuracy.
Note that the standard deviation of the i.i.d. Gaussian noise $\bs n$ is fixed to 0.015.

\bibliography{sample,gmvae}

\begin{thebibliography}{10}
\urlstyle{rm}
\expandafter\ifx\csname url\endcsname\relax
  \def\url#1{\texttt{#1}}\fi
\expandafter\ifx\csname urlprefix\endcsname\relax\def\urlprefix{URL }\fi
\expandafter\ifx\csname doiprefix\endcsname\relax\def\doiprefix{DOI: }\fi
\providecommand{\bibinfo}[2]{#2}
\providecommand{\eprint}[2][]{\url{#2}}

\bibitem{psaltis2016}
\bibinfo{author}{Psaltis, D.} \& \bibinfo{author}{Moser, C.}
\newblock \bibinfo{journal}{\bibinfo{title}{Imaging with multimode fibers}}.
\newblock {\emph{\JournalTitle{Optics and Photonics News}}}
  \textbf{\bibinfo{volume}{27}}, \bibinfo{pages}{24--31}
  (\bibinfo{year}{2016}).

\bibitem{stellinga2021}
\bibinfo{author}{Stellinga, D.} \emph{et~al.}
\newblock \bibinfo{journal}{\bibinfo{title}{Time-of-flight 3d imaging through
  multimode optical fibers}}.
\newblock {\emph{\JournalTitle{Science}}} \textbf{\bibinfo{volume}{374}},
  \bibinfo{pages}{1395--1399} (\bibinfo{year}{2021}).

\bibitem{cizmar2012}
\bibinfo{author}{{\v{C}}i{\v{z}}m{\'a}r, T.} \& \bibinfo{author}{Dholakia, K.}
\newblock \bibinfo{journal}{\bibinfo{title}{Exploiting multimode waveguides for
  pure fibre-based imaging}}.
\newblock {\emph{\JournalTitle{Nature communications}}}
  \textbf{\bibinfo{volume}{3}}, \bibinfo{pages}{1--9} (\bibinfo{year}{2012}).

\bibitem{bianchi2012}
\bibinfo{author}{Bianchi, S.} \& \bibinfo{author}{Di~Leonardo, R.}
\newblock \bibinfo{journal}{\bibinfo{title}{A multi-mode fiber probe for
  holographic micromanipulation and microscopy}}.
\newblock {\emph{\JournalTitle{Lab on a Chip}}} \textbf{\bibinfo{volume}{12}},
  \bibinfo{pages}{635--639} (\bibinfo{year}{2012}).

\bibitem{ploschner2015}
\bibinfo{author}{Pl{\"o}schner, M.}, \bibinfo{author}{Tyc, T.} \&
  \bibinfo{author}{{\v{C}}i{\v{z}}m{\'a}r, T.}
\newblock \bibinfo{journal}{\bibinfo{title}{Seeing through chaos in multimode
  fibres}}.
\newblock {\emph{\JournalTitle{Nature Photonics}}}
  \textbf{\bibinfo{volume}{9}}, \bibinfo{pages}{529--535}
  (\bibinfo{year}{2015}).

\bibitem{loterie2015}
\bibinfo{author}{Loterie, D.} \emph{et~al.}
\newblock \bibinfo{journal}{\bibinfo{title}{Digital confocal microscopy through
  a multimode fiber}}.
\newblock {\emph{\JournalTitle{Optics express}}} \textbf{\bibinfo{volume}{23}},
  \bibinfo{pages}{23845--23858} (\bibinfo{year}{2015}).

\bibitem{caravaca2017}
\bibinfo{author}{Caravaca-Aguirre, A.~M.} \& \bibinfo{author}{Piestun, R.}
\newblock \bibinfo{journal}{\bibinfo{title}{Single multimode fiber endoscope}}.
\newblock {\emph{\JournalTitle{Optics express}}} \textbf{\bibinfo{volume}{25}},
  \bibinfo{pages}{1656--1665} (\bibinfo{year}{2017}).

\bibitem{popoff2010}
\bibinfo{author}{Popoff, S.~M.} \emph{et~al.}
\newblock \bibinfo{journal}{\bibinfo{title}{Measuring the transmission matrix
  in optics: an approach to the study and control of light propagation in
  disordered media}}.
\newblock {\emph{\JournalTitle{Physical review letters}}}
  \textbf{\bibinfo{volume}{104}}, \bibinfo{pages}{100601}
  (\bibinfo{year}{2010}).

\bibitem{cizmar2011}
\bibinfo{author}{{\v{C}}i{\v{z}}m{\'a}r, T.} \& \bibinfo{author}{Dholakia, K.}
\newblock \bibinfo{journal}{\bibinfo{title}{Shaping the light transmission
  through a multimode optical fibre: complex transformation analysis and
  applications in biophotonics}}.
\newblock {\emph{\JournalTitle{Optics express}}} \textbf{\bibinfo{volume}{19}},
  \bibinfo{pages}{18871--18884} (\bibinfo{year}{2011}).

\bibitem{carpenter2014}
\bibinfo{author}{Carpenter, J.}, \bibinfo{author}{Eggleton, B.~J.} \&
  \bibinfo{author}{Schr{\"o}der, J.}
\newblock \bibinfo{journal}{\bibinfo{title}{110x110 optical mode transfer
  matrix inversion}}.
\newblock {\emph{\JournalTitle{Optics express}}} \textbf{\bibinfo{volume}{22}},
  \bibinfo{pages}{96--101} (\bibinfo{year}{2014}).

\bibitem{n2018}
\bibinfo{author}{N’Gom, M.}, \bibinfo{author}{Norris, T.~B.},
  \bibinfo{author}{Michielssen, E.} \& \bibinfo{author}{Nadakuditi, R.~R.}
\newblock \bibinfo{journal}{\bibinfo{title}{Mode control in a multimode fiber
  through acquiring its transmission matrix from a reference-less optical
  system}}.
\newblock {\emph{\JournalTitle{Optics letters}}} \textbf{\bibinfo{volume}{43}},
  \bibinfo{pages}{419--422} (\bibinfo{year}{2018}).

\bibitem{li2021}
\bibinfo{author}{Li, S.} \emph{et~al.}
\newblock \bibinfo{journal}{\bibinfo{title}{Compressively sampling the optical
  transmission matrix of a multimode fibre}}.
\newblock {\emph{\JournalTitle{Light: Science \& Applications}}}
  \textbf{\bibinfo{volume}{10}}, \bibinfo{pages}{1--15} (\bibinfo{year}{2021}).

\bibitem{choi2012}
\bibinfo{author}{Choi, Y.} \emph{et~al.}
\newblock \bibinfo{journal}{\bibinfo{title}{Scanner-free and wide-field
  endoscopic imaging by using a single multimode optical fiber}}.
\newblock {\emph{\JournalTitle{Physical review letters}}}
  \textbf{\bibinfo{volume}{109}}, \bibinfo{pages}{203901}
  (\bibinfo{year}{2012}).

\bibitem{amitonova2018}
\bibinfo{author}{Amitonova, L.~V.} \& \bibinfo{author}{De~Boer, J.~F.}
\newblock \bibinfo{journal}{\bibinfo{title}{Compressive imaging through a
  multimode fiber}}.
\newblock {\emph{\JournalTitle{Optics letters}}} \textbf{\bibinfo{volume}{43}},
  \bibinfo{pages}{5427--5430} (\bibinfo{year}{2018}).

\bibitem{caravaca2019}
\bibinfo{author}{Caravaca-Aguirre, A.~M.} \emph{et~al.}
\newblock \bibinfo{journal}{\bibinfo{title}{Hybrid photoacoustic-fluorescence
  microendoscopy through a multimode fiber using speckle illumination}}.
\newblock {\emph{\JournalTitle{Apl Photonics}}} \textbf{\bibinfo{volume}{4}},
  \bibinfo{pages}{096103} (\bibinfo{year}{2019}).

\bibitem{lan2019}
\bibinfo{author}{Lan, M.} \emph{et~al.}
\newblock \bibinfo{journal}{\bibinfo{title}{Robust compressive multimode fiber
  imaging against bending with enhanced depth of field}}.
\newblock {\emph{\JournalTitle{Optics Express}}} \textbf{\bibinfo{volume}{27}},
  \bibinfo{pages}{12957--12962} (\bibinfo{year}{2019}).

\bibitem{lan2020}
\bibinfo{author}{Lan, M.} \emph{et~al.}
\newblock \bibinfo{journal}{\bibinfo{title}{Averaging speckle patterns to
  improve the robustness of compressive multimode fiber imaging against fiber
  bend}}.
\newblock {\emph{\JournalTitle{Optics Express}}} \textbf{\bibinfo{volume}{28}},
  \bibinfo{pages}{13662--13669} (\bibinfo{year}{2020}).

\bibitem{caravaca2013}
\bibinfo{author}{Caravaca-Aguirre, A.~M.}, \bibinfo{author}{Niv, E.},
  \bibinfo{author}{Conkey, D.~B.} \& \bibinfo{author}{Piestun, R.}
\newblock \bibinfo{journal}{\bibinfo{title}{Real-time resilient focusing
  through a bending multimode fiber}}.
\newblock {\emph{\JournalTitle{Optics express}}} \textbf{\bibinfo{volume}{21}},
  \bibinfo{pages}{12881--12887} (\bibinfo{year}{2013}).

\bibitem{farahi2013}
\bibinfo{author}{Farahi, S.}, \bibinfo{author}{Ziegler, D.},
  \bibinfo{author}{Papadopoulos, I.~N.}, \bibinfo{author}{Psaltis, D.} \&
  \bibinfo{author}{Moser, C.}
\newblock \bibinfo{journal}{\bibinfo{title}{Dynamic bending compensation while
  focusing through a multimode fiber}}.
\newblock {\emph{\JournalTitle{Optics express}}} \textbf{\bibinfo{volume}{21}},
  \bibinfo{pages}{22504--22514} (\bibinfo{year}{2013}).

\bibitem{gu2015}
\bibinfo{author}{Gu, R.~Y.}, \bibinfo{author}{Mahalati, R.~N.} \&
  \bibinfo{author}{Kahn, J.~M.}
\newblock \bibinfo{journal}{\bibinfo{title}{Design of flexible multi-mode fiber
  endoscope}}.
\newblock {\emph{\JournalTitle{Optics express}}} \textbf{\bibinfo{volume}{23}},
  \bibinfo{pages}{26905--26918} (\bibinfo{year}{2015}).

\bibitem{li2021memory}
\bibinfo{author}{Li, S.}, \bibinfo{author}{Horsley, S.~A.},
  \bibinfo{author}{Tyc, T.}, \bibinfo{author}{{\v{C}}i{\v{z}}m{\'a}r, T.} \&
  \bibinfo{author}{Phillips, D.~B.}
\newblock \bibinfo{journal}{\bibinfo{title}{Memory effect assisted imaging
  through multimode optical fibres}}.
\newblock {\emph{\JournalTitle{Nature Communications}}}
  \textbf{\bibinfo{volume}{12}}, \bibinfo{pages}{1--13} (\bibinfo{year}{2021}).

\bibitem{moran2018}
\bibinfo{author}{Moran, O.}, \bibinfo{author}{Caramazza, P.},
  \bibinfo{author}{Faccio, D.} \& \bibinfo{author}{Murray-Smith, R.}
\newblock \bibinfo{journal}{\bibinfo{title}{Deep, complex, invertible networks
  for inversion of transmission effects in multimode optical fibres}}.
\newblock {\emph{\JournalTitle{Advances in Neural Information Processing
  Systems}}} \textbf{\bibinfo{volume}{31}} (\bibinfo{year}{2018}).

\bibitem{borhani2018}
\bibinfo{author}{Borhani, N.}, \bibinfo{author}{Kakkava, E.},
  \bibinfo{author}{Moser, C.} \& \bibinfo{author}{Psaltis, D.}
\newblock \bibinfo{journal}{\bibinfo{title}{Learning to see through multimode
  fibers}}.
\newblock {\emph{\JournalTitle{Optica}}} \textbf{\bibinfo{volume}{5}},
  \bibinfo{pages}{960--966} (\bibinfo{year}{2018}).

\bibitem{rahmani2018}
\bibinfo{author}{Rahmani, B.}, \bibinfo{author}{Loterie, D.},
  \bibinfo{author}{Konstantinou, G.}, \bibinfo{author}{Psaltis, D.} \&
  \bibinfo{author}{Moser, C.}
\newblock \bibinfo{journal}{\bibinfo{title}{Multimode optical fiber
  transmission with a deep learning network}}.
\newblock {\emph{\JournalTitle{Light: Science \& Applications}}}
  \textbf{\bibinfo{volume}{7}}, \bibinfo{pages}{1--11} (\bibinfo{year}{2018}).

\bibitem{turpin2018}
\bibinfo{author}{Turpin, A.}, \bibinfo{author}{Vishniakou, I.} \&
  \bibinfo{author}{d~Seelig, J.}
\newblock \bibinfo{journal}{\bibinfo{title}{Light scattering control in
  transmission and reflection with neural networks}}.
\newblock {\emph{\JournalTitle{Optics express}}} \textbf{\bibinfo{volume}{26}},
  \bibinfo{pages}{30911--30929} (\bibinfo{year}{2018}).

\bibitem{kakkava2019}
\bibinfo{author}{Kakkava, E.} \emph{et~al.}
\newblock \bibinfo{journal}{\bibinfo{title}{Imaging through multimode fibers
  using deep learning: The effects of intensity versus holographic recording of
  the speckle pattern}}.
\newblock {\emph{\JournalTitle{Optical Fiber Technology}}}
  \textbf{\bibinfo{volume}{52}}, \bibinfo{pages}{101985}
  (\bibinfo{year}{2019}).

\bibitem{caramazza2019}
\bibinfo{author}{Caramazza, P.}, \bibinfo{author}{Moran, O.},
  \bibinfo{author}{Murray-Smith, R.} \& \bibinfo{author}{Faccio, D.}
\newblock \bibinfo{journal}{\bibinfo{title}{Transmission of natural scene
  images through a multimode fibre}}.
\newblock {\emph{\JournalTitle{Nature communications}}}
  \textbf{\bibinfo{volume}{10}}, \bibinfo{pages}{1--6} (\bibinfo{year}{2019}).

\bibitem{fan2019}
\bibinfo{author}{Fan, P.}, \bibinfo{author}{Zhao, T.} \& \bibinfo{author}{Su,
  L.}
\newblock \bibinfo{journal}{\bibinfo{title}{Deep learning the high variability
  and randomness inside multimode fibers}}.
\newblock {\emph{\JournalTitle{Optics express}}} \textbf{\bibinfo{volume}{27}},
  \bibinfo{pages}{20241--20258} (\bibinfo{year}{2019}).

\bibitem{li2020}
\bibinfo{author}{Li, Y.} \emph{et~al.}
\newblock \bibinfo{journal}{\bibinfo{title}{Image reconstruction using
  pre-trained autoencoder on multimode fiber imaging system}}.
\newblock {\emph{\JournalTitle{IEEE Photonics Technology Letters}}}
  \textbf{\bibinfo{volume}{32}}, \bibinfo{pages}{779--782}
  (\bibinfo{year}{2020}).

\bibitem{zhao2021}
\bibinfo{author}{Zhao, J.} \emph{et~al.}
\newblock \bibinfo{journal}{\bibinfo{title}{High-fidelity imaging through
  multimode fibers via deep learning}}.
\newblock {\emph{\JournalTitle{Journal of Physics: Photonics}}}
  \textbf{\bibinfo{volume}{3}}, \bibinfo{pages}{015003} (\bibinfo{year}{2021}).

\bibitem{liu2022}
\bibinfo{author}{Liu, Z.} \emph{et~al.}
\newblock \bibinfo{journal}{\bibinfo{title}{All-fiber high-speed image
  detection enabled by deep learning}}.
\newblock {\emph{\JournalTitle{Nature communications}}}
  \textbf{\bibinfo{volume}{13}}, \bibinfo{pages}{1--8} (\bibinfo{year}{2022}).

\bibitem{mitton2022l}
\bibinfo{author}{Mitton, J.} \emph{et~al.}
\newblock \bibinfo{journal}{\bibinfo{title}{Bessel equivariant networks for
  inversion of transmission effects in multi-mode optical fibres}}.
\newblock {\emph{\JournalTitle{arXiv preprint arXiv:2207.12849}}}
  (\bibinfo{year}{2022}).

\bibitem{li2018}
\bibinfo{author}{Li, Y.}, \bibinfo{author}{Xue, Y.} \& \bibinfo{author}{Tian,
  L.}
\newblock \bibinfo{journal}{\bibinfo{title}{Deep speckle correlation: a deep
  learning approach toward scalable imaging through scattering media}}.
\newblock {\emph{\JournalTitle{Optica}}} \textbf{\bibinfo{volume}{5}},
  \bibinfo{pages}{1181--1190} (\bibinfo{year}{2018}).

\bibitem{li2021displacement}
\bibinfo{author}{Li, Y.}, \bibinfo{author}{Cheng, S.}, \bibinfo{author}{Xue,
  Y.} \& \bibinfo{author}{Tian, L.}
\newblock \bibinfo{journal}{\bibinfo{title}{Displacement-agnostic coherent
  imaging through scatter with an interpretable deep neural network}}.
\newblock {\emph{\JournalTitle{Optics Express}}} \textbf{\bibinfo{volume}{29}},
  \bibinfo{pages}{2244--2257} (\bibinfo{year}{2021}).

\bibitem{starshynov2022}
\bibinfo{author}{Starshynov, I.}, \bibinfo{author}{Turpin, A.},
  \bibinfo{author}{Binner, P.} \& \bibinfo{author}{Faccio, D.}
\newblock \bibinfo{journal}{\bibinfo{title}{Statistical dependencies beyond
  linear correlations in light scattered by disordered media}}.
\newblock {\emph{\JournalTitle{Physical Review Research}}}
  \textbf{\bibinfo{volume}{4}}, \bibinfo{pages}{L022033}
  (\bibinfo{year}{2022}).

\bibitem{resisi2021}
\bibinfo{author}{Resisi, S.}, \bibinfo{author}{Popoff, S.~M.} \&
  \bibinfo{author}{Bromberg, Y.}
\newblock \bibinfo{journal}{\bibinfo{title}{Image transmission through a
  dynamically perturbed multimode fiber by deep learning}}.
\newblock {\emph{\JournalTitle{Laser \& Photonics Reviews}}}
  \textbf{\bibinfo{volume}{15}}, \bibinfo{pages}{2000553}
  (\bibinfo{year}{2021}).

\bibitem{kingma2014}
\bibinfo{author}{Kingma, D.} \& \bibinfo{author}{Welling, M.}
\newblock \bibinfo{title}{Auto-encoding variational bayes}.
\newblock In \emph{\bibinfo{booktitle}{ICLR}} (\bibinfo{year}{2014}).

\bibitem{kingma2019}
\bibinfo{author}{Kingma, D.~P.} \& \bibinfo{author}{Welling, M.}
\newblock \bibinfo{journal}{\bibinfo{title}{An introduction to variational
  autoencoders}}.
\newblock {\emph{\JournalTitle{Foundations and Trends in Machine Learning}}}
  \textbf{\bibinfo{volume}{12}}, \bibinfo{pages}{307--392}
  (\bibinfo{year}{2019}).

\bibitem{tonolini2020}
\bibinfo{author}{Tonolini, F.}, \bibinfo{author}{Radford, J.},
  \bibinfo{author}{Turpin, A.}, \bibinfo{author}{Faccio, D.} \&
  \bibinfo{author}{Murray-Smith, R.}
\newblock \bibinfo{journal}{\bibinfo{title}{Variational inference for
  computational imaging inverse problems}}.
\newblock {\emph{\JournalTitle{The Journal of Machine Learning Research}}}
  \textbf{\bibinfo{volume}{21}}, \bibinfo{pages}{7285--7330}
  (\bibinfo{year}{2020}).

\bibitem{shu2016}
\bibinfo{author}{Shu, R.}
\newblock \bibinfo{title}{Gaussian mixture vae: Lessons in variational
  inference, generative models, and deep nets}.

\bibitem{figueroa2019}
\bibinfo{author}{Figueroa, J.~A.}
\newblock \bibinfo{title}{Semi-supervised learning using deep generative models
  and auxiliary tasks}.
\newblock In \emph{\bibinfo{booktitle}{NeurIPS}} (\bibinfo{year}{2019}).

\bibitem{collier2019}
\bibinfo{author}{Collier, M.} \& \bibinfo{author}{Urdiales, H.}
\newblock \bibinfo{journal}{\bibinfo{title}{Scalable deep unsupervised
  clustering with concrete gmvaes}}.
\newblock {\emph{\JournalTitle{arXiv preprint arXiv:1909.08994}}}
  (\bibinfo{year}{2019}).

\bibitem{charakorn2020}
\bibinfo{author}{Charakorn, R.} \emph{et~al.}
\newblock \bibinfo{journal}{\bibinfo{title}{An explicit local and global
  representation disentanglement framework with applications in deep clustering
  and unsupervised object detection}}.
\newblock {\emph{\JournalTitle{arXiv preprint arXiv:2001.08957}}}
  (\bibinfo{year}{2020}).

\bibitem{varolgunecs2020}
\bibinfo{author}{Varolg{\"u}ne{\c{s}}, Y.~B.}, \bibinfo{author}{Bereau, T.} \&
  \bibinfo{author}{Rudzinski, J.~F.}
\newblock \bibinfo{journal}{\bibinfo{title}{Interpretable embeddings from
  molecular simulations using gaussian mixture variational autoencoders}}.
\newblock {\emph{\JournalTitle{Machine Learning: Science and Technology}}}
  \textbf{\bibinfo{volume}{1}}, \bibinfo{pages}{015012} (\bibinfo{year}{2020}).

\bibitem{franceschi2020}
\bibinfo{author}{Franceschi, J.-Y.}, \bibinfo{author}{Delasalles, E.},
  \bibinfo{author}{Chen, M.}, \bibinfo{author}{Lamprier, S.} \&
  \bibinfo{author}{Gallinari, P.}
\newblock \bibinfo{title}{Stochastic latent residual video prediction}.
\newblock In \emph{\bibinfo{booktitle}{International Conference on Machine
  Learning}}, \bibinfo{pages}{3233--3246} (\bibinfo{organization}{PMLR},
  \bibinfo{year}{2020}).

\bibitem{wu2021}
\bibinfo{author}{Wu, B.}, \bibinfo{author}{Nair, S.},
  \bibinfo{author}{Martin-Martin, R.}, \bibinfo{author}{Fei-Fei, L.} \&
  \bibinfo{author}{Finn, C.}
\newblock \bibinfo{title}{Greedy hierarchical variational autoencoders for
  large-scale video prediction}.
\newblock In \emph{\bibinfo{booktitle}{Proceedings of the IEEE/CVF Conference
  on Computer Vision and Pattern Recognition}}, \bibinfo{pages}{2318--2328}
  (\bibinfo{year}{2021}).

\bibitem{wang2022}
\bibinfo{author}{Wang, L.}, \bibinfo{author}{Tan, H.}, \bibinfo{author}{Zhou,
  F.}, \bibinfo{author}{Zuo, W.} \& \bibinfo{author}{Sun, P.}
\newblock \bibinfo{journal}{\bibinfo{title}{Unsupervised anomaly video
  detection via a double-flow convlstm variational autoencoder}}.
\newblock {\emph{\JournalTitle{IEEE Access}}} \textbf{\bibinfo{volume}{10}},
  \bibinfo{pages}{44278--44289} (\bibinfo{year}{2022}).

\end{thebibliography}


\begin{thebibliography}{10}
\urlstyle{rm}
\expandafter\ifx\csname url\endcsname\relax
  \def\url#1{\texttt{#1}}\fi
\expandafter\ifx\csname urlprefix\endcsname\relax\def\urlprefix{URL }\fi
\expandafter\ifx\csname doiprefix\endcsname\relax\def\doiprefix{DOI: }\fi
\providecommand{\bibinfo}[2]{#2}
\providecommand{\eprint}[2][]{\url{#2}}

\bibitem{jang2016}
\bibinfo{author}{Jang, E.}, \bibinfo{author}{Gu, S.} \& \bibinfo{author}{Poole,
  B.}
\newblock \bibinfo{journal}{\bibinfo{title}{Categorical reparameterization with
  gumbel-softmax}}.
\newblock {\emph{\JournalTitle{arXiv preprint arXiv:1611.01144}}}
  (\bibinfo{year}{2016}).

\bibitem{maddison2017}
\bibinfo{author}{Maddison, C.}, \bibinfo{author}{Mnih, A.} \&
  \bibinfo{author}{Teh, Y.}
\newblock \bibinfo{title}{The concrete distribution: A continuous relaxation of
  discrete random variables}.
\newblock In \emph{\bibinfo{booktitle}{ICLR}} (\bibinfo{year}{2017}).

\bibitem{figueroa2019}
\bibinfo{author}{Figueroa, J.~A.}
\newblock \bibinfo{title}{Semi-supervised learning using deep generative models
  and auxiliary tasks}.
\newblock In \emph{\bibinfo{booktitle}{NeurIPS}} (\bibinfo{year}{2019}).

\bibitem{collier2019}
\bibinfo{author}{Collier, M.} \& \bibinfo{author}{Urdiales, H.}
\newblock \bibinfo{journal}{\bibinfo{title}{Scalable deep unsupervised
  clustering with concrete gmvaes}}.
\newblock {\emph{\JournalTitle{arXiv preprint arXiv:1909.08994}}}
  (\bibinfo{year}{2019}).

\bibitem{weinberger2006}
\bibinfo{author}{Weinberger, K.~Q.}, \bibinfo{author}{Blitzer, J.} \&
  \bibinfo{author}{Saul, L.~K.}
\newblock \bibinfo{title}{Distance metric learning for large margin nearest
  neighbor classification}.
\newblock In \emph{\bibinfo{booktitle}{Adv. Neural Inf. Process. Syst.}},
  \bibinfo{pages}{1473--1480} (\bibinfo{year}{2006}).

\bibitem{schroff2015}
\bibinfo{author}{Schroff, F.}, \bibinfo{author}{Kalenichenko, D.} \&
  \bibinfo{author}{Philbin, J.}
\newblock \bibinfo{title}{Facenet: A unified embedding for face recognition and
  clustering}.
\newblock In \emph{\bibinfo{booktitle}{IEEE CVPR}}, \bibinfo{pages}{815--823}
  (\bibinfo{year}{2015}).

\bibitem{kingma2015}
\bibinfo{author}{Kingma, D.~P.} \& \bibinfo{author}{Ba, J.~L.}
\newblock \bibinfo{title}{Adam: A method for stochastic gradient descent}.
\newblock In \emph{\bibinfo{booktitle}{ICLR}}, \bibinfo{pages}{1--15}
  (\bibinfo{year}{2015}).

\bibitem{li2018}
\bibinfo{author}{Li, Y.}, \bibinfo{author}{Xue, Y.} \& \bibinfo{author}{Tian,
  L.}
\newblock \bibinfo{journal}{\bibinfo{title}{Deep speckle correlation: a deep
  learning approach toward scalable imaging through scattering media}}.
\newblock {\emph{\JournalTitle{Optica}}} \textbf{\bibinfo{volume}{5}},
  \bibinfo{pages}{1181--1190} (\bibinfo{year}{2018}).

\bibitem{li2021displacement}
\bibinfo{author}{Li, Y.}, \bibinfo{author}{Cheng, S.}, \bibinfo{author}{Xue,
  Y.} \& \bibinfo{author}{Tian, L.}
\newblock \bibinfo{journal}{\bibinfo{title}{Displacement-agnostic coherent
  imaging through scatter with an interpretable deep neural network}}.
\newblock {\emph{\JournalTitle{Optics Express}}} \textbf{\bibinfo{volume}{29}},
  \bibinfo{pages}{2244--2257} (\bibinfo{year}{2021}).

\bibitem{resisi2021}
\bibinfo{author}{Resisi, S.}, \bibinfo{author}{Popoff, S.~M.} \&
  \bibinfo{author}{Bromberg, Y.}
\newblock \bibinfo{journal}{\bibinfo{title}{Image transmission through a
  dynamically perturbed multimode fiber by deep learning}}.
\newblock {\emph{\JournalTitle{Laser \& Photonics Reviews}}}
  \textbf{\bibinfo{volume}{15}}, \bibinfo{pages}{2000553}
  (\bibinfo{year}{2021}).

\bibitem{stellinga2021}
\bibinfo{author}{Stellinga, D.} \emph{et~al.}
\newblock \bibinfo{journal}{\bibinfo{title}{Time-of-flight 3d imaging through
  multimode optical fibers}}.
\newblock {\emph{\JournalTitle{Science}}} \textbf{\bibinfo{volume}{374}},
  \bibinfo{pages}{1395--1399} (\bibinfo{year}{2021}).

\end{thebibliography}

\section*{Acknowledgements}
\label{sec:supp:acknowledge}
This work was supported by the Royal Academy of Engineering under the Research Fellowship scheme RF201617/16/31 and by the Engineering and Physical Sciences Research Council (EPSRC) Grant number EP/T00097X/1.

\section*{Data availability}

The datasets used and/or analyzed during the current study available from the corresponding author on reasonable request.

\section*{Author contributions statement}
All authors (A.A., S.P.M., Y.A., M.J.P., and S.M.) contributed to the the methodology and the writing of the manuscript. A.A. and S.P.M. conducted the experiments, tests and investigation. Y.A., M.J.P., and S.M. provided supervision throughout the project.

\section*{Competing interests}

The authors declare no competing interests.

\end{document}


\flushbottom
\maketitle

\section*{Supplementary Note 1: The experimental geometry}
\label{sec:supp1}
We use a Q-switched, 21.7 kHz repetition rate, 532 nm wavelength, 700 ps, pulsed laser (Teem Photonics SNG-100P-1x0) for all imaging procedures in this work. The beam is expanded such that the 1/e$^2$ diameter fits inside the short axis of a DMD (Vialux V-7000). A constant grating is applied to the DMD to direct diffracted light away from the 0$^{\text{th}}$ order where phase shaping may be applied. The beam is then directed through a half wave plate (HWP) which is used to control the power incident on the fiber and direct a small portion of the light to a triggering photodiode (PD) through a polarizing beam splitter (PBS) as shown in Supplementary Figure~\ref{fig:setup_supp}. The light is then converted into circularly polarized light with a quarter wave plate (QWP). In a plane Fourier to the DMD, the light is spatially filtered to select the 1$^{\text{st}}$ order of the diffraction pattern. The beam is then collimated with an objective lens (Olympus 40x Plan-N) and the plane wave incident on the fiber is coupled into the fiber core. This makes the fiber input facet conjugate to the plane of the DMD. For brevity, extra relay lenses are excluded from Supplementary Figure~\ref{fig:setup_supp}, however the magnification from the fiber to the DMD is set at 1600:9. The illumination fiber is a 50 cm, graded index, 0.275 NA, 62.5~$\mu$m core diameter, fiber (Thorlabs GIF625). It is epoxied into a 20 G, 3 cm, needle alongside a 50 cm, step index, 0.39 NA, 400~$\mu$m core diameter, collection fiber (Thorlabs FT400UMT). Both fibers are housed in a common jacket and are therefore both subject to bending by the bending arm and rotation stage.

During calibration, an internal reference beam is used to measure the phase of the probed fiber modes. The reference beam is made by randomly selecting 26 points at a plane Fourier to the DMD and exciting these points using a super position of grating frequencies on the DMD. A probe mode is then added to the super position and phase stepped through the reference. The resulting interference patterns are incident on a white screen and are recorded by a camera (Hamamatsu Orca Flash 4.0). From these images, the complex transmission matrix of the fiber can be determined, allowing for the calculation of the optimal DMD phase mask needed to generate any spot within the fiber's illumination cone.

For imaging, the camera is not used. Rather, an object is placed at the screen and the calculated DMD phase masks are displayed on the DMD at the rate of one pattern per laser pulse. The trigger PD starts a set number of samples being recorded by the APD (Menlo Systems APD210) at 2.5 GS/s as the pulse propagates to the object and back. The reflectivity of the object is given by the peak intensity of the return pulse and, as such, the maximum of the return signal is recorded giving a single value for each scanned spot. This time gating increases the signal to noise ratio of the system. As the system is numerically trained and physically tested, great care was taken to ensure that the perceived resolution of the images being recorded by the fiber was identical to that used in the training.

Calibration was performed once in this work at an extremum of the fiber bend apparatus and, as such, most of the imaging was performed with unknown speckle patterns. Reconstructing such images with unknown and unoptimized speckle patterns was the task borne out by the GMVAE. To ensure the expected speckle patterns differed significantly from the calibrated configuration, the bend arm and the rotation stage worked in opposition to one another as seen in Supplementary Figure~\ref{fig:setup_supp}, that is, while one pulls the fiber in one direction the other pushes another part of the fiber in the opposing direction. The bend arm had a range of movement of 8 cm, and the rotation stage turned the needle, camera, and screen together through 50$^\circ$. At each of the 11 stops, the speckle patterns were recorded using the same DMD phase masks calculated during the calibration in the case of the calibrated experiment, and the same randomly selected DMD phase masks in the uncalibrated experiment. A selection of these were then used in simulating the training data for the GMVAE.
\begin{figure}[h!]%
\centering
\includegraphics[width=1\textwidth]{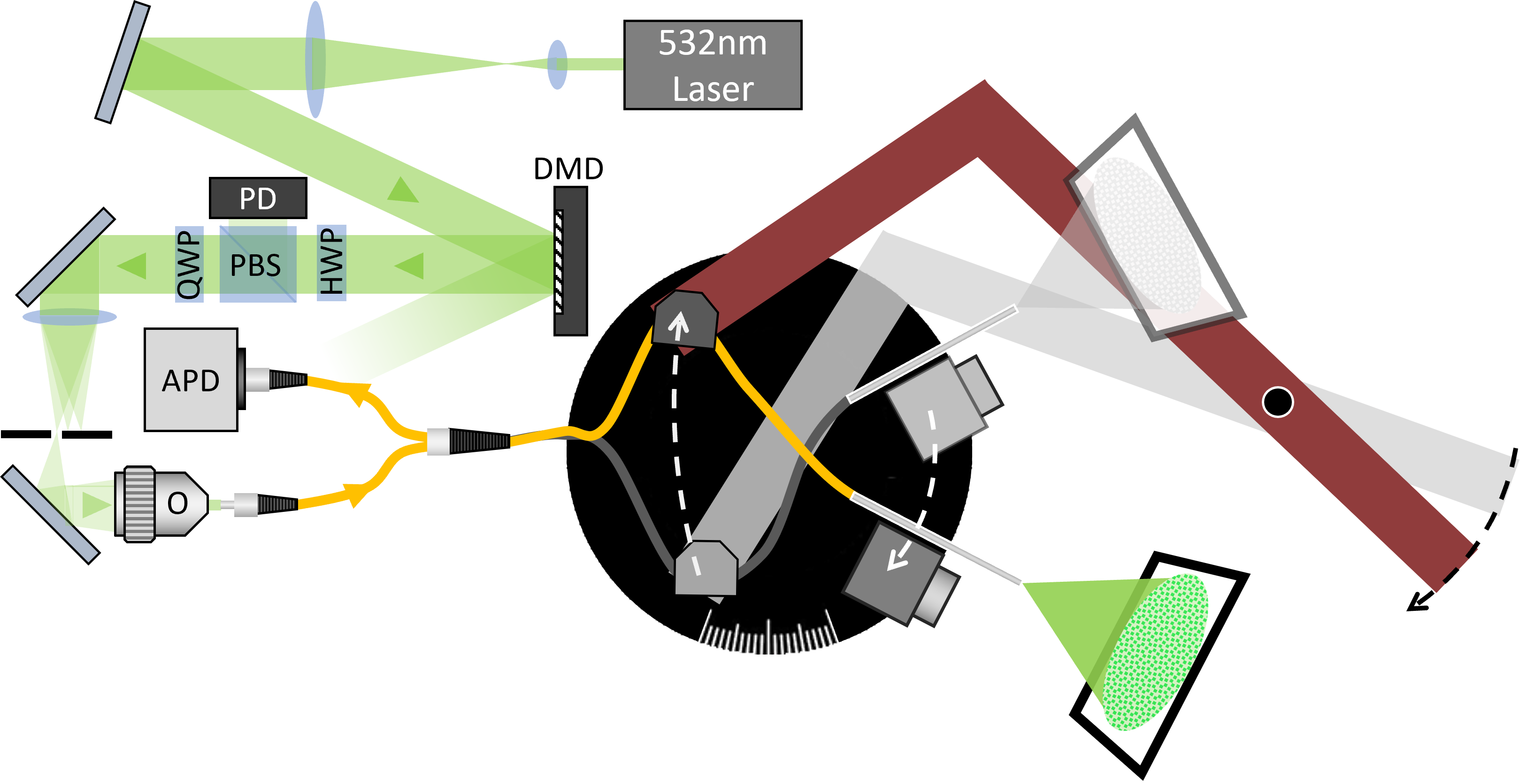}
\caption{Schematic illustrating the calibration, imaging, and bending apparatus used in this work. The grayed-out camera, screen, needle, and bend arm represent the fiber configuration at the starting position where calibration is performed, while the corresponding fully saturated portion depicts the fiber configuration at the final position. \textbf{DMD}: digital micromirror device; \textbf{HWP}: half-wave plate; \textbf{PBS}: polarizing beam splitter; \textbf{QWP}: quarter-wave plate; \textbf{PD}: photodiode; \textbf{O}: objective lens; \textbf{APD}: avalanche photodiode.}\label{fig:setup_supp}
\end{figure}

\section*{Supplementary Note 2: Gaussian mixture variational autoencoder}\label{sec:supp2}
Recall that our aim is to learn a configuration-agnostic variational autoencoder that can estimate the underlying object $\bs{x}$ and its class from APD measurements $\bs{y}$ corresponding to a bend that was not part of the training set. The class of each image can be either represented by a scalar $c$ or a one-hot encoded vector $\bs{c}$ of length $k$, i.e., the number of classes. In GMVAE, the image to be reconstructed $\bs{x}$ is assumed to follow a generative process:
%
\begin{subequations}
\begin{align}
\label{eq:1a}
\mathrm{p}_{\bs{\theta}}(\bs{x},  \bs{z}, \bs{c}) &= \mathrm{p}_{\bs{\theta}}(\bs{x} \vert \bs{z}) ~\mathrm{p}_{\bs{\theta}}(\bs{z} \vert \bs{c}) ~\mathrm{p}(\bs{c}), \\
\label{eq:1b}
c &\sim \text{{Cat}}(c \vert ~1/k), \\
\label{eq:1c}
\bs{z} \vert \bs{c} &\sim \mathcal{N}(\bs{z} \vert  ~{\bs{\mu}}_{{\bs{z}}_{\bs{\theta}}}(\bs{c}),  {\bs{\sigma}}^2_{{\bs{z}}_{\bs{\theta}}}(\bs{c})), \\
\label{eq:1d}
\bs{x} \vert \bs{z} &\sim \mathcal{N}(\bs{x} \vert ~{\bs{\mu}}_{{\bs{x}}_{\bs{\theta}}}(\bs{z}), {\bs{\sigma}}^2_{{\bs{x}}_{\bs{\theta}}}(\bs{z})),
\end{align}
\end{subequations}
%
where $\bs{z} \in \mathbb{R}^d$ is the latent variable. The vectors ${\bs{\mu}}_{{\bs{z}}_{\bs{\theta}}}$,  ${\bs{\sigma}}^2_{{\bs{z}}_{\bs{\theta}}}$,  ${\bs{\mu}}_{{\bs{x}}_{\bs{\theta}}}$ and ${\bs{\sigma}}^2_{{\bs{x}}_{\bs{\theta}}}$ are given by the GMVAE generative network with parameters ${\bs{\theta}}$. 
The objective of GMVAE is to approximate the posterior distribution $\mathrm{p}(\bs{z},  \bs{c} \vert \bs{y})$ which is intractable and typically approximated by a factorized posterior,  namely the inference model:
%
\begin{subequations}
\begin{align}
\label{eq:1a}
\mathrm{q}_{\bs{\phi}}(\bs{z},  \bs{c} \vert \bs{y}) &= \mathrm{q}_{\bs{\phi}}(\bs{z}\vert \bs{y},   \bs{c}) ~\mathrm{q}_{\bs{\phi}}(\bs{c} \vert \bs{y}), \\
\label{eq:1b}
c \vert \bs{y} &\sim \text{{Cat}}(c \vert ~\pi_{\bs{\phi}}(\bs{y})), \\
\label{eq:1c}
\bs{z} \vert \bs{y}, \bs{c} &\sim \mathcal{N}(\bs{z} \vert  ~{\bs{\mu}}_{{\bs{z}_{\bs{\phi}}}}(\bs{y},  \bs{c}),  {\bs{\sigma}}^2_{{\bs{z}_{\bs{\phi}}}}(\bs{y},  \bs{c})),
\end{align}
\end{subequations}
%
where ${\bs{\phi}}$ are the parameters of the inference network. 
The categorical distribution $\text{{Cat}}(c \vert ~\pi_{\bs{\phi}}(\bs{y}))$ is approximated using the Gumbel-Softmax (Concrete) distribution~\cite{jang2016,  maddison2017}.
The parameters of the inference and generative networks, ${\bs{\phi}}$ and ${\bs{\theta}}$, are learnt by maximizing the evidence lower-bound (ELBO)~\cite{figueroa2019, collier2019},  that is a lower bound on the log probability of the observations:
%
\begin{equation}
\label{eq:elbo}
\text{ELBO} = - \alpha \text{KL} (\mathrm{q}_{\bs{\phi}}(\bs{z}\vert \bs{y},   \bs{c}) \| ~\mathrm{p}_{\bs{\theta}}(\bs{z} \vert \bs{c}) ) 
- \beta \text{KL} (\mathrm{q}_{\bs{\phi}}(\bs{c} \vert \bs{y}) \| ~\mathrm{p}(\bs{c}))
+  \omega \mathrm{E}_{\mathrm{q}_{\bs{\phi}}(\bs{z}\vert \bs{y},   \bs{c})} [\log \mathrm{p}_{\bs{\theta}}(\bs{x} \vert \bs{z})],
\end{equation}
%
where $\alpha, \beta, \omega > 0$ are hyper-parameters controlling the weights of the different priors. KL stands for the Kullback-Leibler divergence and the last term is given by the reconstruction loss.

The classification accuracy can be greatly enhanced by learning an appropriate distance metric from the labels.
Therefore, we adopt the triplet embedding loss as an auxiliary loss function to regularize the latent space~\cite{weinberger2006, schroff2015}. 
The triplet loss penalizes large distances between features belonging to the same class while penalizing small distances between features with non-matching labels. The final objective for the adopted GMVAE takes the form:
\begin{equation}
\label{eq:9}
\mathcal{L}_{\text{GMVAE}} = \text{ELBO} + \gamma \mathcal{L}_{\text{triplet}},
\end{equation}
 where $\mathcal{L}_{\text{triplet}}$ is the triplet embedding loss and $\gamma > 0$ is the hyper-parameter controlling the weight of the auxiliary loss function.  
Using grid search, we found that $\alpha=1, ~\beta=200,~\omega=50$ and $\gamma=50$ gives optimal results. The same parameters are fixed for all experiments.

\section*{Supplementary Note 3: GMVAE architecture}
\label{sec:supp3}
The GMVAE architecture consists of inference and generative networks.
The inference network starts with an encoder to learn the features of the input measurement vector $\bs{y}$, dubbed $\hat{\bs{y}}$.
The encoder consists of three convolutional layers each followed by a batch normalization layer and a rectified linear unit (ReLU) as an activation function.
The classification network $\mathrm{q}_{\bs{\phi}}(\bs{c} \vert \bs{y})$ consists of two dense layers of size 1024 and $k$ (here $k=8$ classes) and a dropout layer of rate 0.2 between the dense layers to avoid over-fitting.  
The network input is the learnt features $\hat{\bs{y}}$ and the output is the sampled categories given by the Gumbel-Softmax (Concrete) distribution~\cite{jang2016,  maddison2017}.
The inference network $\mathrm{q}_{\bs{\phi}}(\bs{z} \vert \bs{y}, \bs{c})$ has two dense layers of size $d=1024$ to produce the latent space mean ${\bs{\mu}}_{{\bs{z}}_{\bs{\phi}}}$ and variance ${\bs{\sigma}}^2_{{\bs{z}}_{\bs{\phi}}}$.
The inference network input is the concatenation of $\hat{\bs{y}}$ and the sampled categories and the output is the sampled latent space given by the reparameterization trick. 
Note that a dropout of rate 0.2 is applied on the features $\hat{\bs{y}}$ before concatenation with the categories to avoid features over-fitting. 
The prior network $\mathrm{p}_{\bs{\theta}}(\bs{z} \vert \bs{c})$ is represented by two dense layers of size $d=1024$ to produce the prior mean ${\bs{\mu}}_{{\bs{z}}_{\bs{\theta}}}$ and variance ${\bs{\sigma}}^2_{{\bs{z}}_{\bs{\theta}}}$, respectively.
Finally, the generative network $\mathrm{p}_{\bs{\theta}}(\bs{x} \vert \bs{z})$ consists of a dense layer of size 4096 followed by three transposed convolutional layers (each followed by a batch normalization layer and a ReLU activation function) and finally a dense layer of size $N=4096$ with a logistic (sigmoid) activation function to produce the output vector $\bs{\mu}_{{\bs{x}}_{\bs{\theta}}}$. GMVAE was trained over 300 epochs using the Adam optimizer~\cite{kingma2015} with a learning rate $10^{-4}$ and a batch size 100.

\section*{Supplementary Note 4: AE and C-AE architecture}
\label{sec:supp4}

We recall that the performance of the GMVAE is compared to that of the autoencoder (AE)~\cite{li2018,li2021displacement}. 
The structure of the AE follows the encoder–decoder architecture where the input measurements of size $64\times64$ go through the encoder which consists of three dense blocks connected by max pooling layers for downsampling. 
Each dense block contains multiple convolutional layers, each followed by a batch normalization layer and a ReLU activation function.
Next, the downsampled map of the measurements goes through the decoder that consists of three dense blocks connected by upsampling layers.
Similarly to~\cite{li2018,resisi2021} and as opposed to~\cite{li2021displacement}, we do not use fully-connected layers in the bottleneck of the AE as we notice that this leads to degradation in the reconstruction quality.

The classification accuracy of GMVAE is compared to that of the C-AE classifier.
C-AE is trained on reconstructed images from AE and can only be trained after AE is trained while GMVAE is trained on measurements for the classification and reconstruction tasks simultaneously.
C-AE has the same architecture as the classifier used in our GMVAE. It consists of three convolutional layers (each followed by a batch normalization layer and a ReLU activation function) followed by two dense layers of size 1024 and eight (number of classes) with a dropout layer of rate 0.2 between the dense layers to avoid over-fitting.  

AE and C-AE were trained over 100 epochs using the Adam optimizer~\cite{kingma2015} with a learning rate $10^{-4}$ and a batch size 100.

\section*{Supplementary Note 5: Further results from the first experiment with wavefront shaping}
\label{sec:supp5}

\begin{figure}[htb!]
\centering
\begin{subfigure}{0.33\textwidth}
\centering
\small{\textbf{(A)} 1\textsuperscript{st} experiment, $\text{C}_{10}$}\\
\includegraphics[width=1\linewidth]{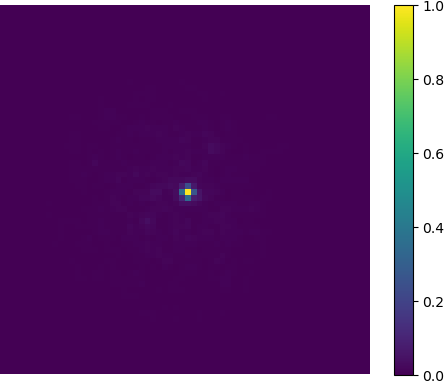}\\

\small{\textbf{(D)} 2\textsuperscript{nd} experiment, $\text{C}_{10}$}\\
\includegraphics[width=1\linewidth]{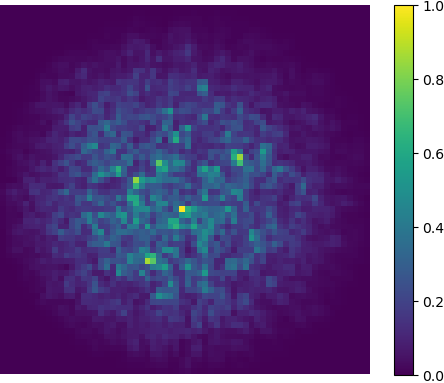}\\
\end{subfigure}%
~
\begin{subfigure}{0.33\textwidth}
\centering
\small{\textbf{(B)} 1\textsuperscript{st} experiment, $\text{C}_{5}$}\\
\includegraphics[width=1\linewidth]{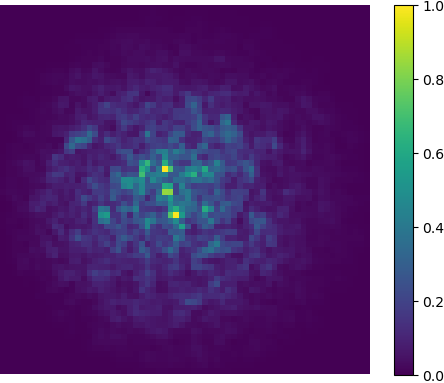}\\

\small{\textbf{(E)} 2\textsuperscript{nd} experiment, $\text{C}_{5}$}\\
\includegraphics[width=1\linewidth]{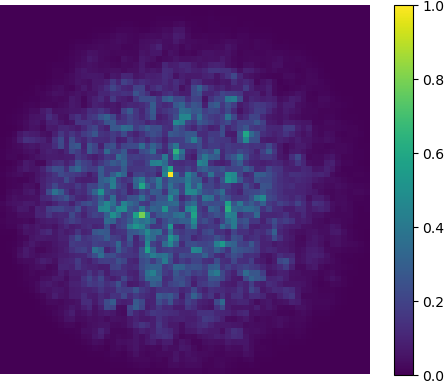}\\
\end{subfigure}%
~
\begin{subfigure}{0.33\textwidth}
\centering
\small{\textbf{(C)} 1\textsuperscript{st} experiment, $\text{C}_{0}$}\\
\includegraphics[width=1\linewidth]{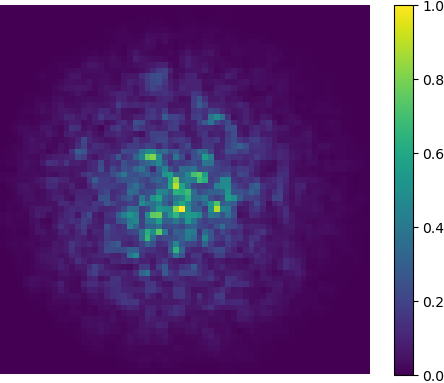}\\

\small{\textbf{(F)} 2\textsuperscript{nd} experiment, $\text{C}_{0}$}\\
\includegraphics[width=1\linewidth]{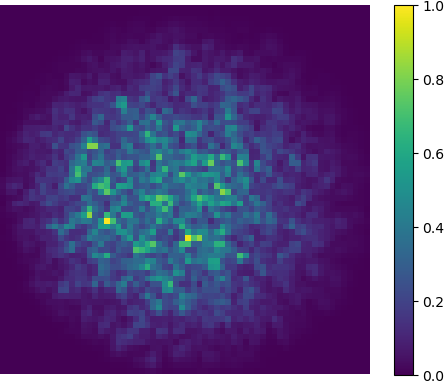}\\
\end{subfigure}%
~
\caption{Speckles from a fixed DMD phase mask recorded at different fiber configurations: $\text{C}_{10}$, $\text{C}_{5}$, and $\text{C}_{0}$. \textbf{(A)}, \textbf{(B)}, and \textbf{(C)} show speckles from the first experiment with wavefront shaping, while \textbf{(D)}, \textbf{(E)}, and \textbf{(F)} depict speckles from the second experiment without wavefront shaping.}
\label{fig:speckles_raster}
\end{figure}

\begin{figure}[htb!]
\centering
\includegraphics[trim={2.2cm 2.5cm 2.2cm 2.1cm},clip,width=0.75\linewidth]{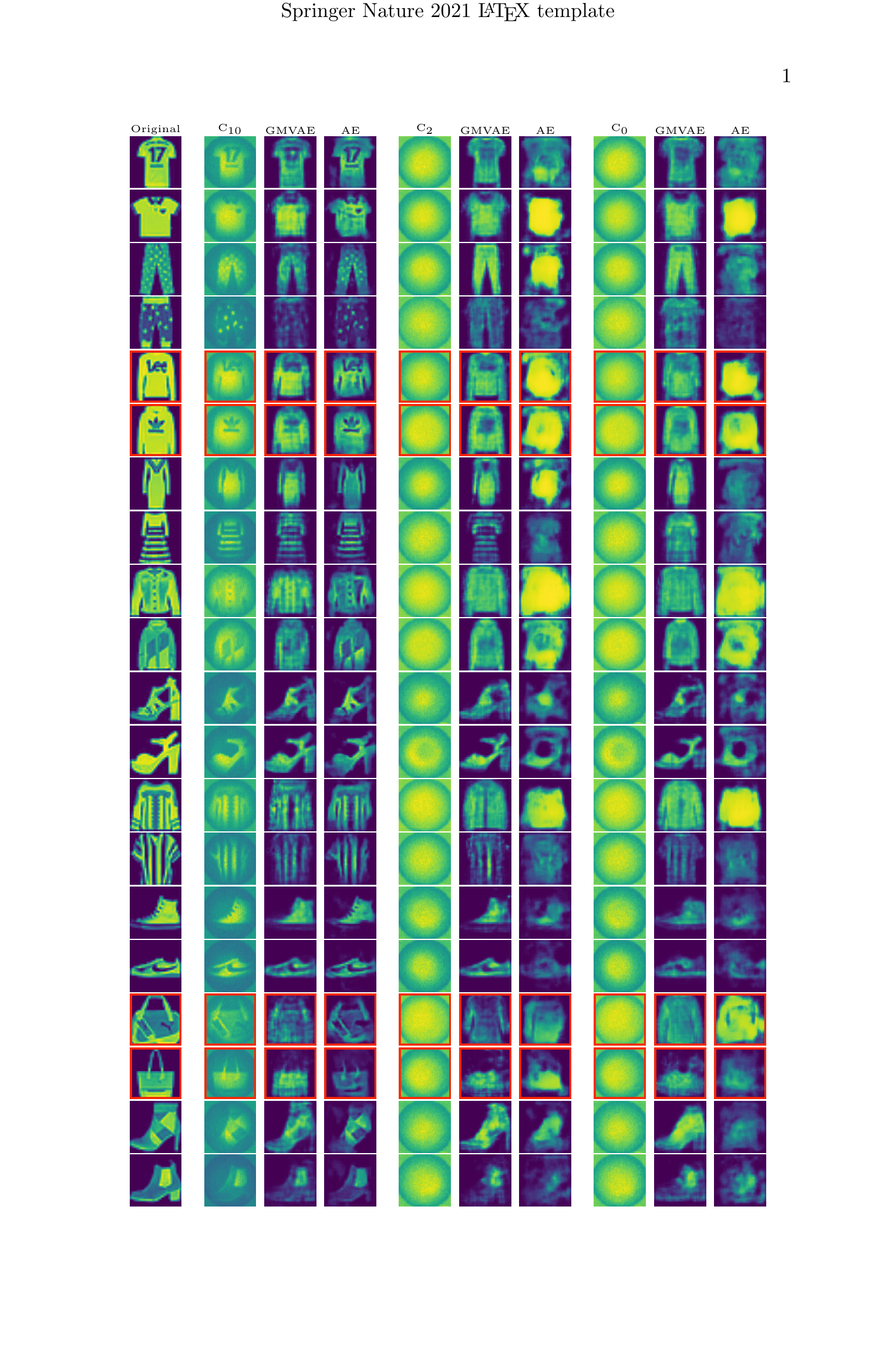}\\
\caption{First experiment with wavefront shaping: reconstruction results at the configurations $\text{C}_{10}$, $\text{C}_{2}$ and $\text{C}_{0}$. The first column displays the original images, while each set of three consecutive columns presents the APD measurements, GMVAE reconstruction, and AE reconstruction. Images highlighted by red squares correspond to new classes not used during training. The intensity values of all images in the figure range between 0 and 1.}
\label{fig:images_raster}
\end{figure}

\begin{figure}[h]
\centering
\centerline{{\textbf{(A)} $\text{C}_{10}$}}
\includegraphics[trim={1.7cm 15.8cm 1.6cm 1.9cm},clip,width=.8\linewidth]{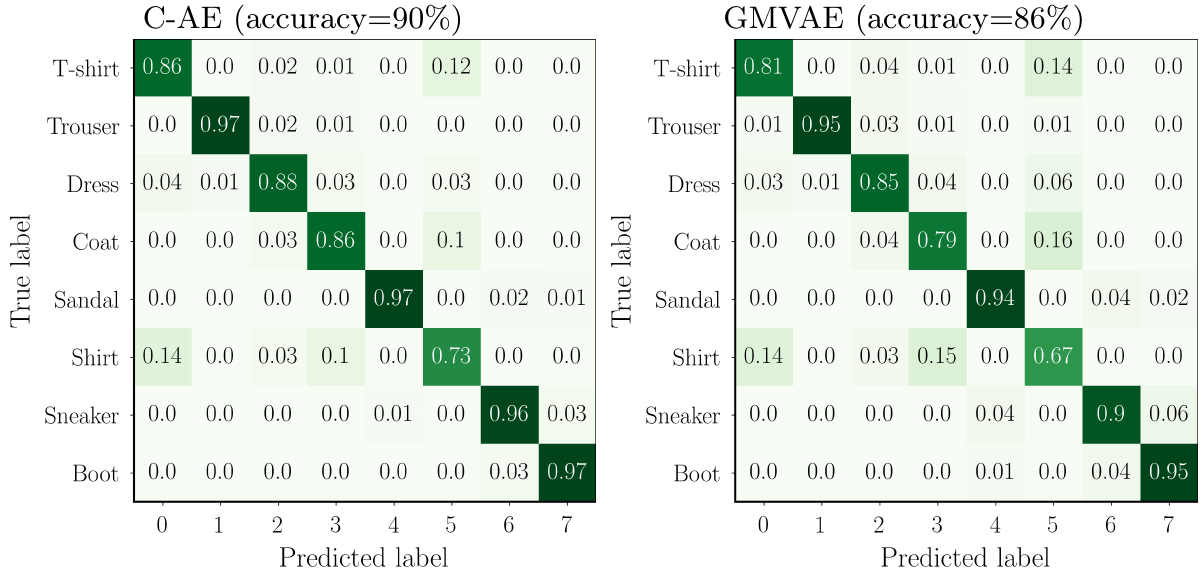}\\
~
\\
\vspace*{-0.2cm}
\centerline{{\textbf{(B)} $\text{C}_{2}$}}
\includegraphics[trim={1.7cm 15.8cm 1.6cm 1.9cm},clip,width=.8\linewidth]{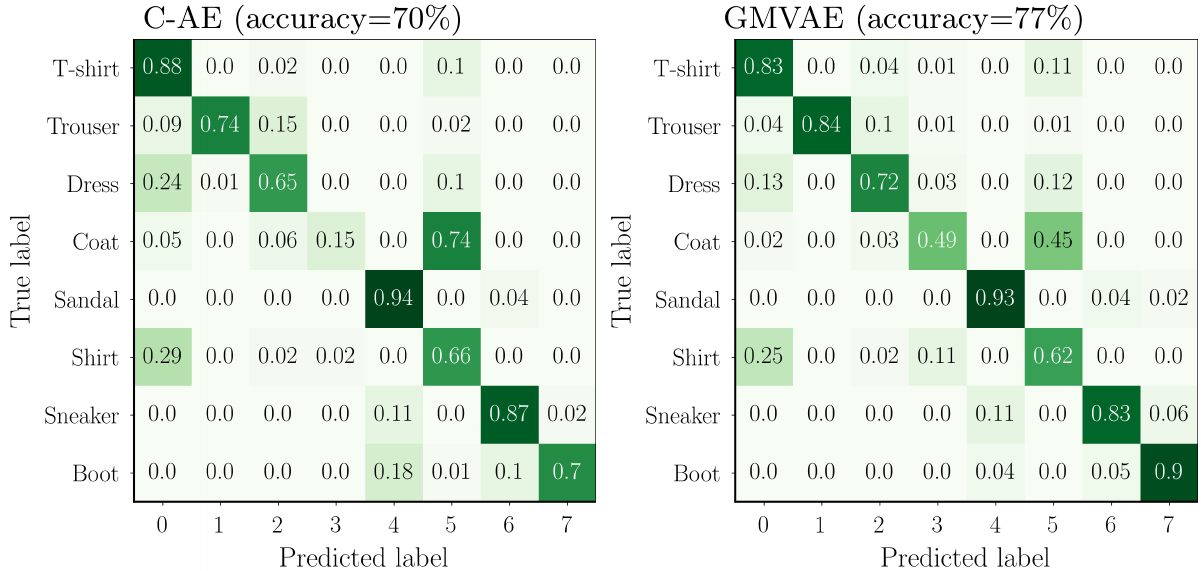}\\
~
\\
\vspace*{-0.2cm}
\centerline{{\textbf{(C)} $\text{C}_{0}$}}
\includegraphics[trim={1.7cm 15.8cm 1.6cm 1.9cm},clip,width=.8\linewidth]{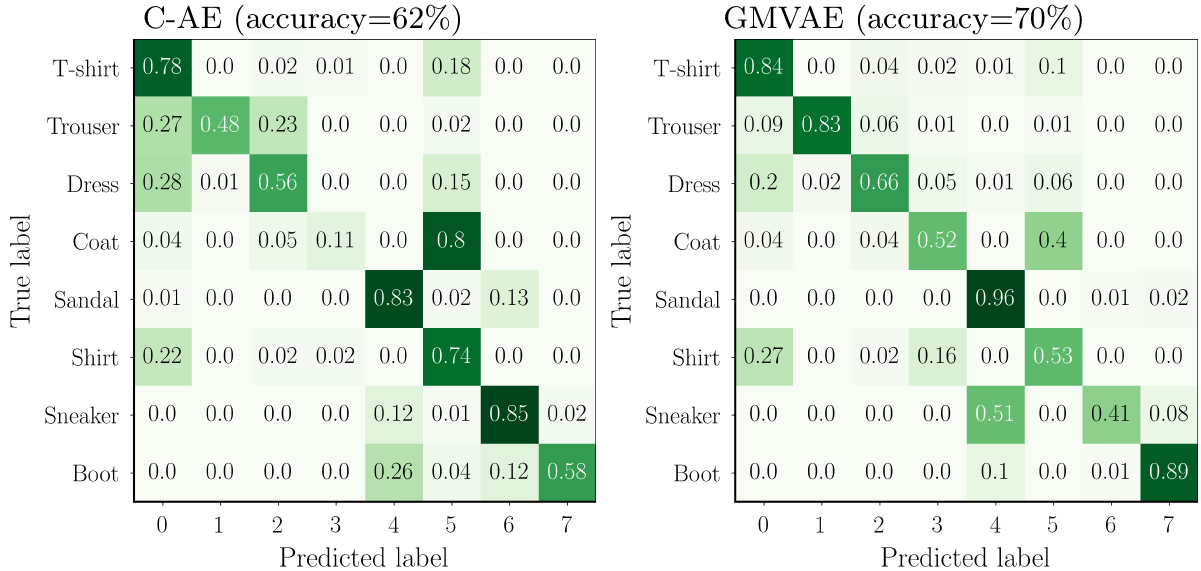}\\
~
\caption{First experiment with wavefront shaping: confusion matrices obtained by C-AE (left) and GMVAE (right) at different configurations \textbf{(A)} $\text{C}_{10}$, \textbf{(B)} $\text{C}_{2}$ and \textbf{(C)} $\text{C}_{0}$ for 8000 numerical measurements from the 8 trained-on classes.}
\label{fig:confusion_raster}
\end{figure}

\begin{figure}[h]
\centering
\centerline{{\textbf{(A)} $\text{C}_{10}$}}
\includegraphics[trim={1.7cm 15.8cm 1.6cm 1.9cm},clip,width=.8\linewidth]{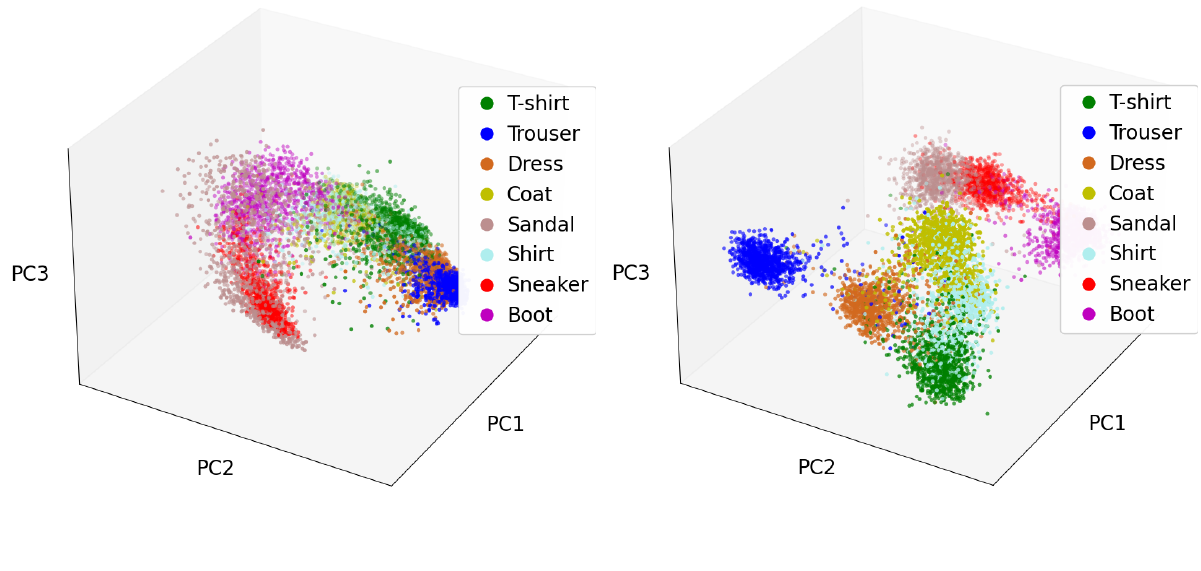}\\
~
\\
\vspace*{-0.3cm}
\centerline{{\textbf{(B)} $\text{C}_{2}$}}
\includegraphics[trim={1.7cm 15.8cm 1.6cm 1.9cm},clip,width=.8\linewidth]{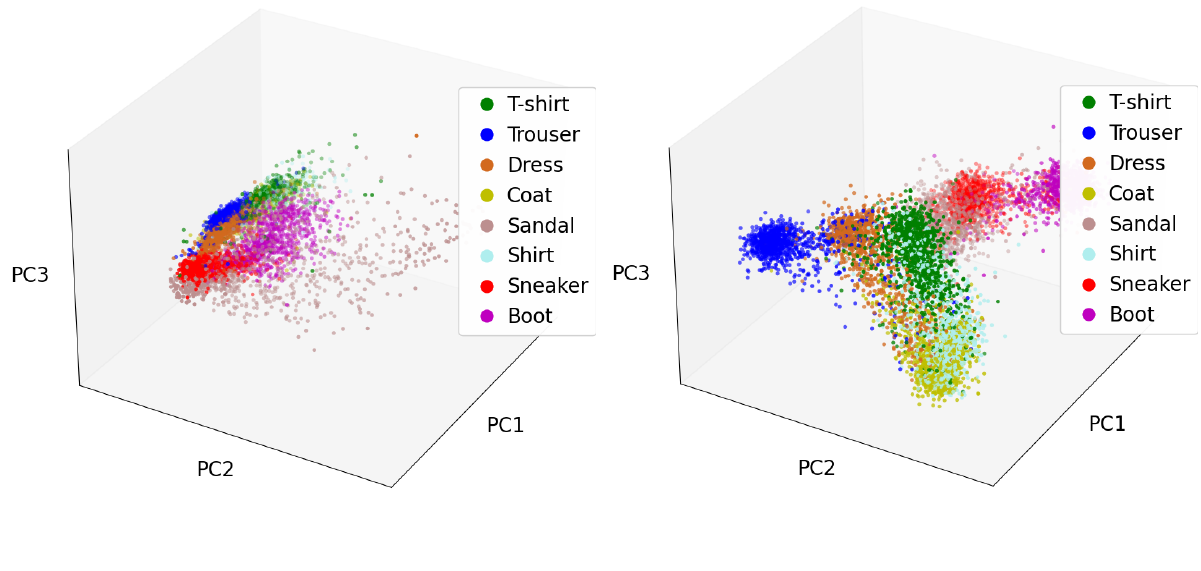}\\
~
\\
\vspace*{-0.3cm}
\centerline{{\textbf{(C)} $\text{C}_{0}$}}
\includegraphics[trim={1.7cm 15.8cm 1.6cm 1.9cm},clip,width=.8\linewidth]{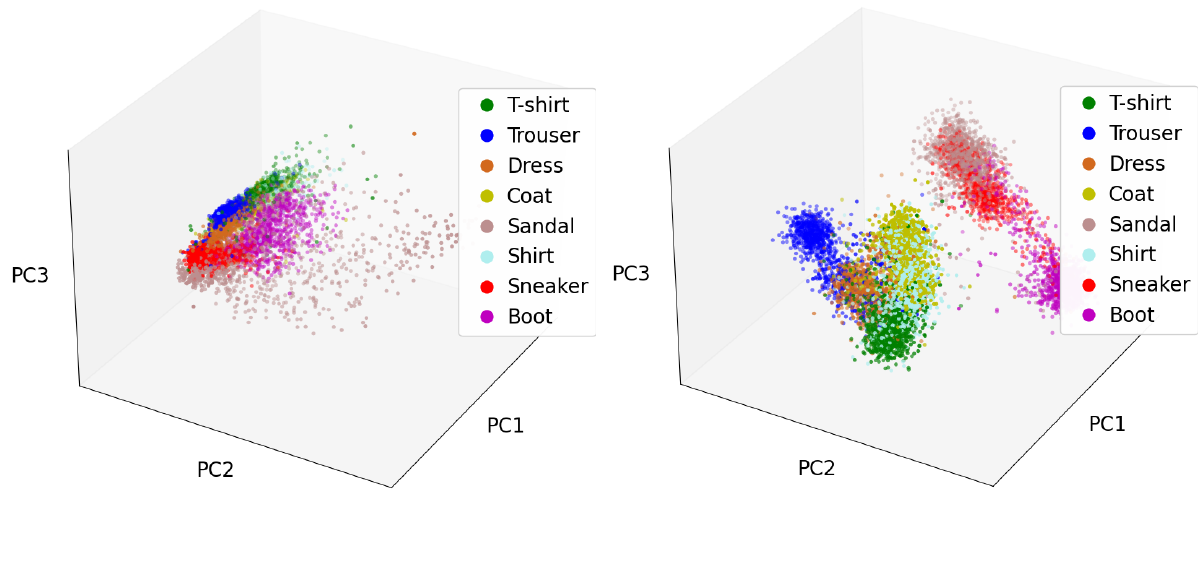}\\
~
\vspace*{-0.3cm}
\caption{First experiment with wavefront shaping: 3D PCA projection of the raw (left) and GMVAE latent space (right) vectors obtained at different configurations \textbf{(A)} $\text{C}_{10}$, \textbf{(B)} $\text{C}_{2}$ and \textbf{(C)} $\text{C}_{0}$ using 8000 numerical measurements from the 8 trained-on classes. }
\label{fig:pca_raster}
\end{figure}

This experiment utilizes the wavefront shaping technique~\cite{stellinga2021} to generate focal spots at the distal end of the fiber. This allows for raster-scanning imaging which can lead to better results due to better SNR ratios. Supplementary Figure~\ref{fig:speckles_raster} (first row) presents speckles from a fixed DMD phase mask recorded at different fiber configurations. 
The speckle at the calibrated configuration $\text{C}_{10}$ represents a simple focal point (Supplementary Figure~\ref{fig:speckles_raster}, \textbf{(A)}). By bending the fiber away from the calibration position, the focal points become speckles due to coupling between spatial modes of the fiber (Supplementary Figure~\ref{fig:speckles_raster}, \textbf{(B)} and \textbf{(C)}).
Supplementary Figure~\ref{fig:images_raster} shows the reconstruction quality of GMVAE and AE on APD measurements collected at $\text{C}_{10}$, $\text{C}_{2}$ and $\text{C}_{0}$.
Both GMVAE and AE give good reconstruction quality on measurements recorded at  $\text{C}_{10}$. This is expected since $\text{C}_{10}$ is the calibrated configuration and was used during training, hence the shapes of the images can be seen directly in the measurements.
However, the results show a superior performance for the GMVAE in comparison to AE for the new configurations $\text{C}_{2}$ and $\text{C}_{0}$ which are far away from the calibration position.
We can see from Supplementary Figure~\ref{fig:confusion_raster}, \textbf{(A)} that C-AE could achieve 90\% classification accuracy with 86\% for GMVAE at the calibrated configuration $\text{C}_{10}$. However,  GMVAE scores 77\% at $\text{C}_{2}$ and 70\% at $\text{C}_{0}$ with 7\% and 8\% enhancement from C-AE, respectively.
Supplementary Figure~\ref{fig:pca_raster} presents the raw data and latent vectors of 8000 numerical measurements computed at the configurations $\text{C}_{10}$, $\text{C}_{2}$ and $\text{C}_{0}$ and projected using principal component analysis (PCA) to the 3D space.
In contrast to the raw data (Supplementary Figure~\ref{fig:pca_raster}, left), the GMVAE latent space (Supplementary Figure~\ref{fig:pca_raster}, right) of the testing data points shows a clear separation between the different classes.

\section*{Supplementary Note 6: Further results from the second experiment without wavefront shaping}
\label{sec:supp6}
\begin{figure}[htb!]
\centering
\includegraphics[trim={2.2cm 2.5cm 2.2cm 2.1cm},clip,width=0.75\linewidth]{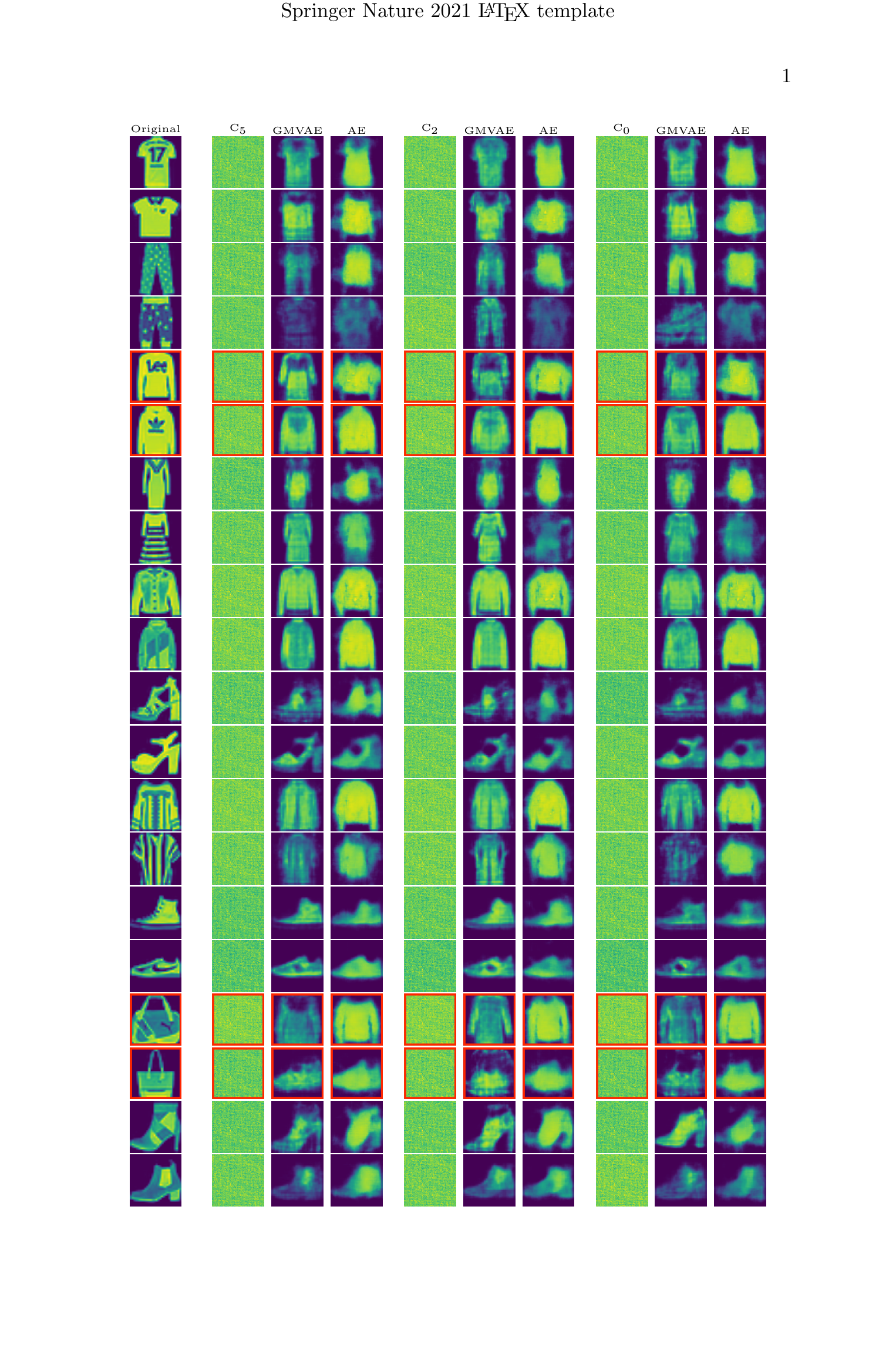}\\
\caption{Second experiment without wavefront shaping: reconstruction results at the configurations $\text{C}_{5}$, $\text{C}_{2}$ and $\text{C}_{0}$. The first column displays the original images, while each set of three consecutive columns presents the APD measurements, GMVAE reconstruction, and AE reconstruction. Images highlighted by red squares correspond to new classes not used during training. The intensity values of all images in the figure range between 0 and 1.}
\label{fig:images_speckle}
\end{figure}

In this experiment, speckles were recorded at the different configurations with no configuration-specific wavefront shaping performed.
Supplementary Figure~\ref{fig:speckles_raster} (second row) shows speckles from a fixed DMD phase mask recorded at different fiber configurations.
We can see that all configurations are equally challenging.
For qualitative assessment, Supplementary Figure~\ref{fig:images_speckle} compares the reconstruction quality of GMVAE against AE on APD measurements collected at the following configurations: $\text{C}_{5}$, $\text{C}_{2}$ and $\text{C}_{0}$.
GMVAE maintained better reconstruction quality compared to AE for all configurations.
However, GMVAE overall reconstruction quality was reduced in comparison with the first experiment with wavefront shaping. This is expected since speckles in the first experiment had a higher signal-to-noise ratio (SNR) than speckles in this experiment because most of the energy was concentrated around the focal points leading to a more orthogonal measurement basis.
The confusion matrices obtained by GMVAE and C-AE at the configurations $\text{C}_{5}$, $\text{C}_{2}$ and $\text{C}_{0}$ on 8000 numerical measurements from the 8 trained-on classes are shown in Supplementary Figure~\ref{fig:confusion_speckle}.     
Once again, the results suggest better classification accuracy for GMVAE compared to C-AE.  More precisely,  GMVAE scores 79\% at $\text{C}_{5}$, 75\% at $\text{C}_{2}$ and 66\% at $\text{C}_{0}$ with 1\%, 7\% and 17\% enhancement from C-AE, respectively.
Finally, the raw data and the GMVAE latent vectors of 8000 numerical measurements computed at the configurations $\text{C}_{5}$, $\text{C}_{2}$ and $\text{C}_{0}$ and projected by PCA to the 3D space are presented in Supplementary Figure~\ref{fig:pca_speckle}.

\section*{Supplementary Movie 1: Video reconstruction of static and moving objects during fiber bend alteration}
\label{sec:video}
The video, provided as supplementary material, presents the reconstruction of three static objects and one moving object as the fiber is bent from the calibrated position $\text{C}_{10}$ (10~mm, 230$^\circ$) to $\text{C}_{5}$ (5~mm, 255$^\circ$). This change corresponds to a fiber bend alteration of 25$^\circ$ and a movement range of 4 cm.

\begin{figure}[h]
\centering
\centerline{{\textbf{(A)} $\text{C}_{5}$}}
\includegraphics[trim={1.7cm 15.8cm 1.6cm 1.9cm},clip,width=.8\linewidth]{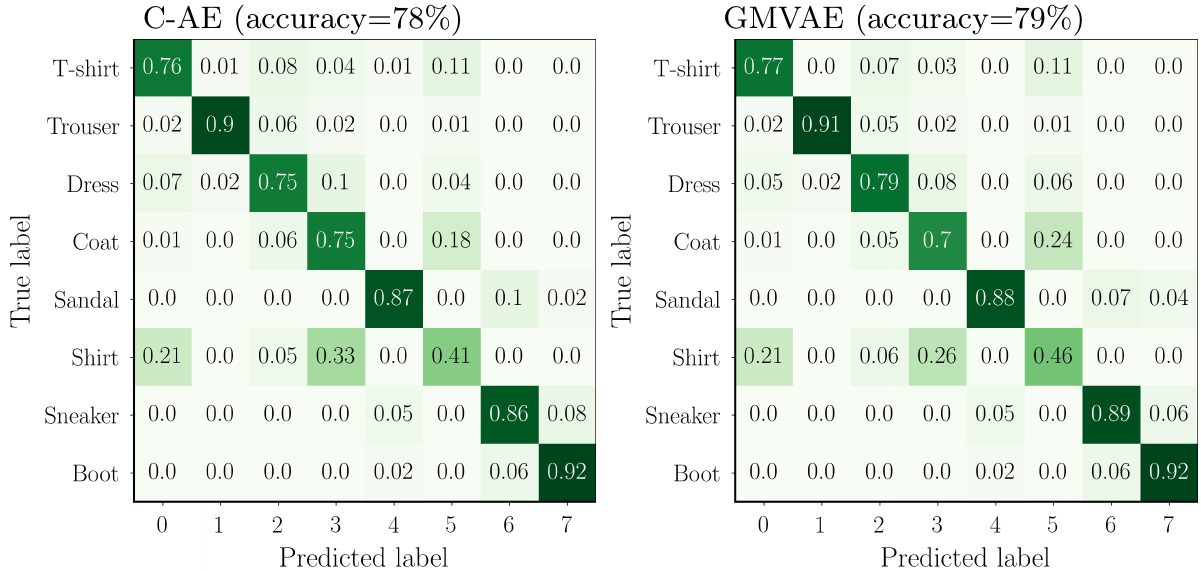}\\
~
\\
\vspace*{-0.2cm}
\centerline{{\textbf{(B)} $\text{C}_{2}$}}
\includegraphics[trim={1.7cm 15.8cm 1.6cm 1.9cm},clip,width=.8\linewidth]{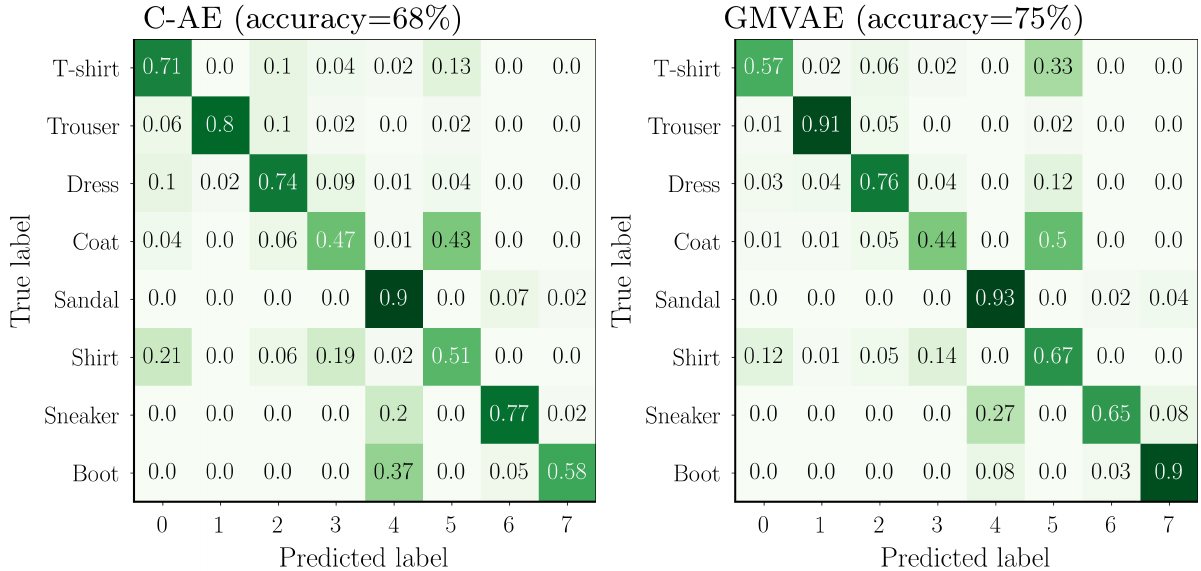}\\
~
\\
\vspace*{-0.2cm}
\centerline{{\textbf{(C)} $\text{C}_{0}$}}
\includegraphics[trim={1.7cm 15.8cm 1.6cm 1.9cm},clip,width=.8\linewidth]{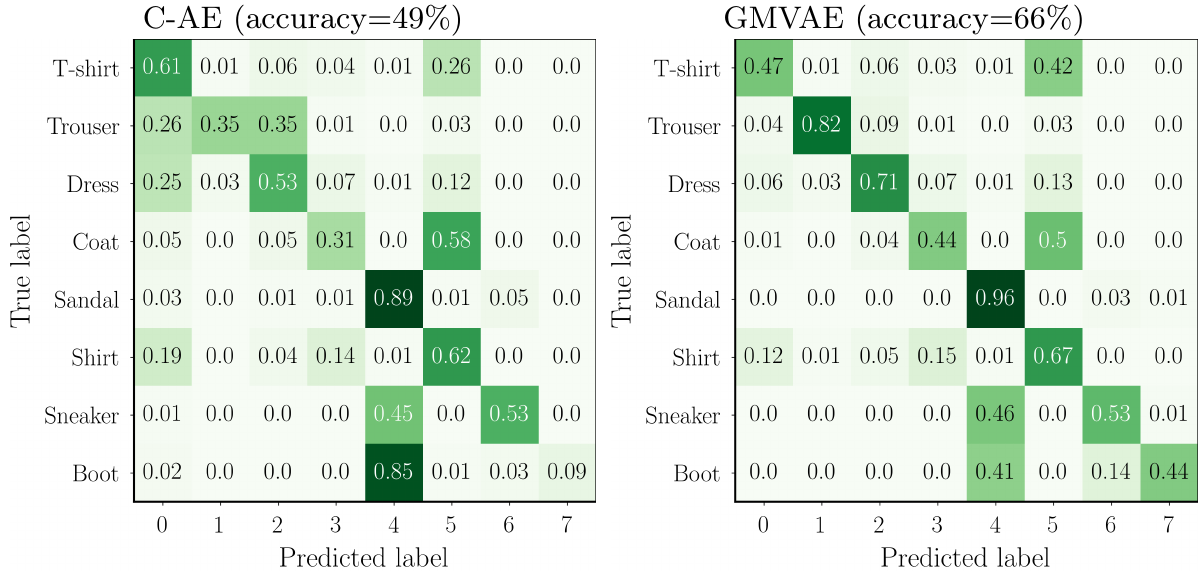}\\
~
\caption{Second experiment without wavefront shaping: confusion matrices obtained by C-AE (left) and GMVAE (right) at different configurations \textbf{(A)} $\text{C}_{5}$, \textbf{(B)} $\text{C}_{2}$ and \textbf{(C)} $\text{C}_{0}$ for 8000 numerical measurements from the 8 trained-on classes.}
\label{fig:confusion_speckle}
\end{figure}

\begin{figure}[h]
\centering
\centerline{{\textbf{(A)} $\text{C}_{5}$}}
\includegraphics[trim={1.7cm 15.8cm 1.6cm 1.9cm},clip,width=.8\linewidth]{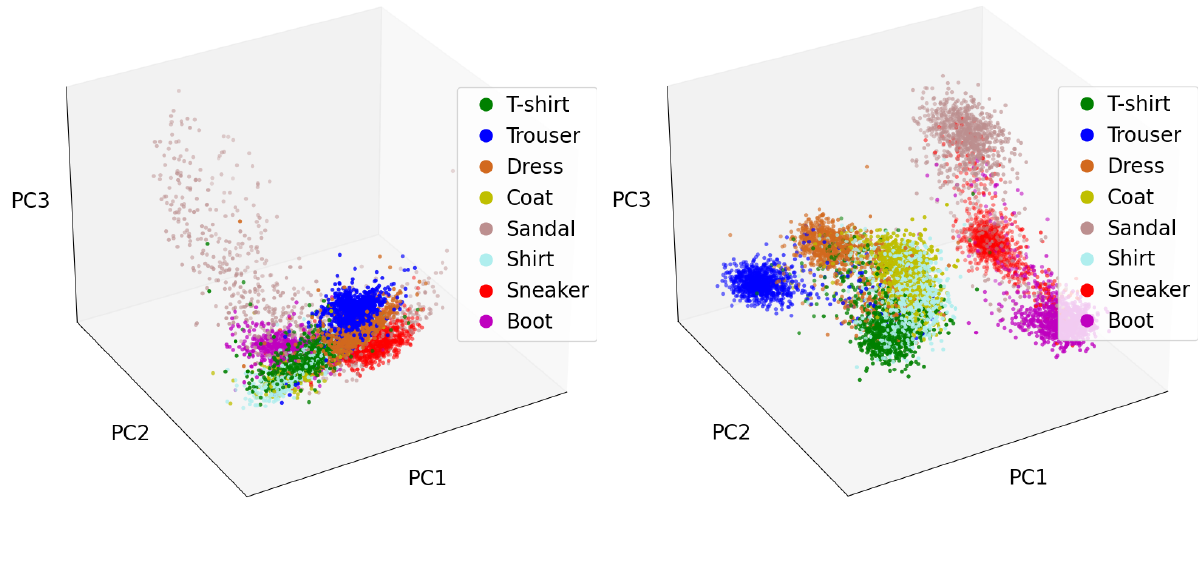}\\
~
\\
\vspace*{-0.3cm}
\centerline{{\textbf{(B)} $\text{C}_{2}$}}
\includegraphics[trim={1.7cm 15.8cm 1.6cm 1.9cm},clip,width=.8\linewidth]{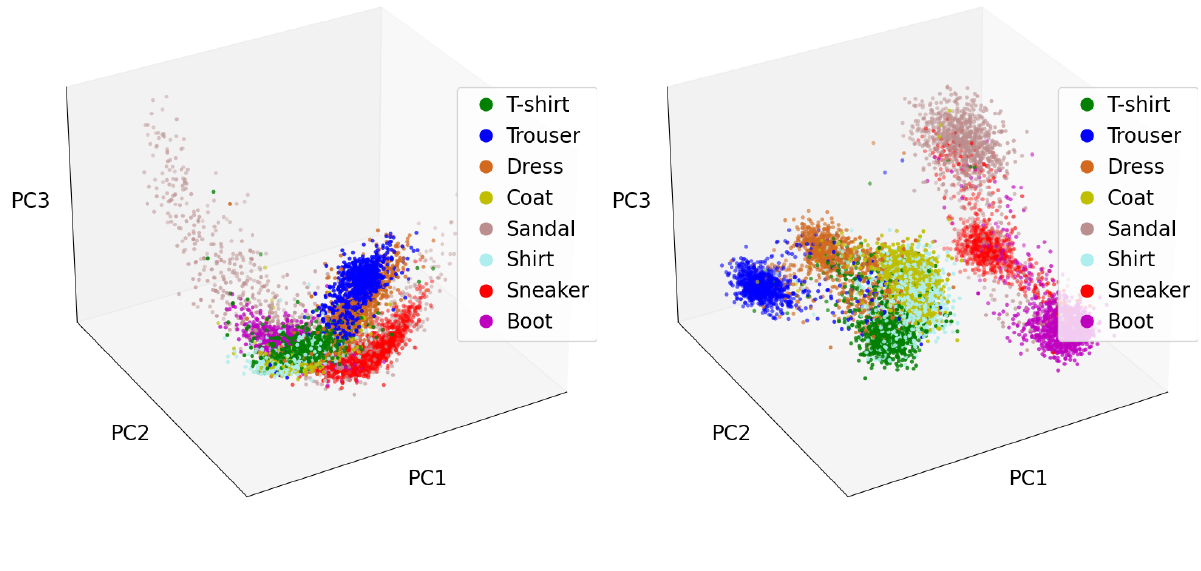}\\
~
\\
\vspace*{-0.3cm}
\centerline{{\textbf{(C)} $\text{C}_{0}$}}
\includegraphics[trim={1.7cm 15.8cm 1.6cm 1.9cm},clip,width=.8\linewidth]{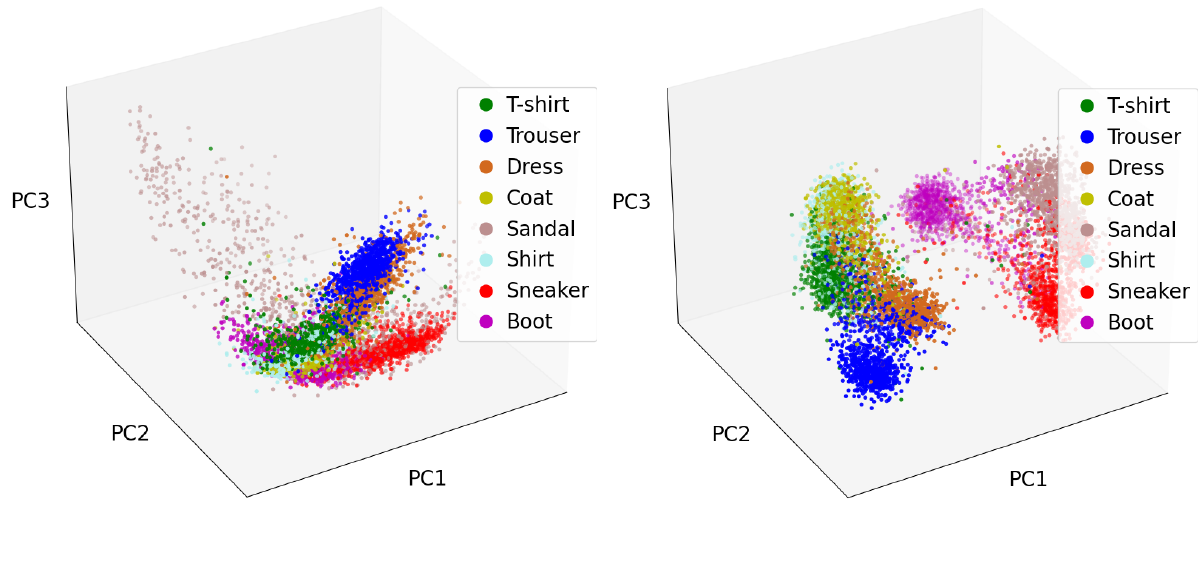}\\
~
\vspace*{-0.3cm}
\caption{Second experiment without wavefront shaping: 3D PCA projection of the raw (left) and GMVAE latent space (right) vectors obtained at different configurations \textbf{(A)} $\text{C}_{5}$, \textbf{(B)} $\text{C}_{2}$ and \textbf{(C)} $\text{C}_{0}$ using 8000 numerical measurements from the 8 trained-on classes. }
\label{fig:pca_speckle}
\end{figure}

\bibliography{sample,gmvae}